\documentclass[11pt]{article}

\usepackage[preprint]{acl}
\usepackage{multirow}
\usepackage{booktabs}
\usepackage{longtable}
\usepackage{amssymb}
\usepackage{array}
\usepackage{amsmath}
\usepackage{times}
\usepackage{latexsym}
\usepackage{subcaption}
\usepackage{algorithm}
\usepackage{algpseudocode}
\usepackage{float}

\usepackage{graphicx}
\usepackage{makecell}
\usepackage[table]{xcolor}

\usepackage{wrapfig}
\usepackage{xcolor}         % colors
\usepackage{multirow}
\usepackage{graphicx}
\usepackage{subcaption}
\usepackage{enumitem}
\usepackage{tcolorbox}
\usepackage{color-edits}
\usepackage{tocloft}
\usepackage{booktabs}
\usepackage{adjustbox}
\usepackage{amsmath}
\usepackage{wasysym}
\newlistof{appfigures}{afg}{List of Comparision Figures in Different Settings}

\extrafloats{100}

\usepackage[normalem]{ulem}
\useunder{\uline}{\ul}{}

\addauthor{gn}{magenta}
\addauthor{ysu}{cyan}
\definecolor{prompt_purple}{RGB}{160, 32, 240}
\definecolor{prompt_green}{RGB}{46, 204, 113}
\definecolor{prompt_blue}{RGB}{41, 128, 185}
\definecolor{prompt_red}{RGB}{192, 57, 43}

\newtcolorbox{promptbox}[2][Prompt]{
colback=black!5!white,
arc=5pt, 
boxrule=0.5pt,
fonttitle=\bfseries,
title=#1, 
before upper={\small}, fontupper=\fontfamily{ptm}\selectfont,
colframe=#2,
}

\usepackage{etoc}
\usepackage{titletoc}

\newcommand\DoToC{%
  \startcontents
  \printcontents{}{1}{\noindent \textbf{\Large{Table of Contents in Appendix}}\vskip3pt\vskip5pt}
  \vskip3pt\vskip5pt
}

\usepackage{pifont}

\newcommand{\greencheck}{\textcolor{green!60!black}{\ding{52}}} 
\newcommand{\redmark}{\textcolor{red}{\ding{55}}}

\usepackage[T1]{fontenc}
\usepackage[utf8]{inputenc}

\usepackage{microtype}

\usepackage{inconsolata}

\usepackage{graphicx}

\title{SWE-QA-Pro: A Representative Benchmark and Scalable Training Recipe for Repository-Level Code Understanding}

\author{
  \textbf{Songcheng Cai}$^{1}$\thanks{Equal contribution}  \quad
  \textbf{Zhiheng Lyu}$^{1}$\footnotemark[1] \quad
  \textbf{Yuansheng Ni}$^{1}$ \quad
  \textbf{Xiangchao Chen}$^{1}$ \\
  \textbf{Baichuan Zhou}$^{1}$ \quad
  \textbf{Shenzhe Zhu}$^{2}$ \quad
  \textbf{Yi Lu}$^{2}$ \quad
  \textbf{Haozhe Wang}$^{3}$ \quad
  \textbf{Chi Ruan}$^{1}$ \\
  \textbf{Benjamin Schneider}$^{1}$ \quad
  \textbf{Weixu Zhang}$^{4}$ \quad
  \textbf{Xiang Li}$^{1}$ \quad
  \textbf{Andy Zheng}$^{1}$ \\
  \textbf{Yuyu Zhang}$^{5}$ \quad
  \textbf{Ping Nie}$^{1}$ \quad
  \textbf{Wenhu Chen}$^{1}$ \thanks{Corresponding author.}
  \\
  $^{1}$University of Waterloo \quad
  $^{2}$University of Toronto \\
  $^{3}$The Hong Kong University of Science and Technology \\
  $^{4}$McGill University \& MILA \quad
  $^{5}$Verdent AI, Inc. \\
\texttt{\href{https://github.com/TIGER-AI-Lab/SWE-QA-Pro}{https://github.com/TIGER-AI-Lab/SWE-QA-Pro}}
}

\begin{document}
\maketitle
\begin{abstract}
% Large language models (LLMs) are increasingly deployed for software engineering tasks that require reasoning over entire repositories rather than isolated code snippets. However, existing code and repository-level QA benchmarks provide limited support for evaluating and training such capabilities, often focus on a small set of popular projects and include many questions that can be resolved without substantive engagement with the codebase, instead drawing on generic knowledge already captured during pretraining.
% We present SWE-QA-Pro, a repository-level QA benchmark that targets these limitations. SWE-QA-Pro is built from long-tail repositories with executable environments, enforces topical diversity via issue-driven clustering, and systematically filters out questions solvable by strong direct-answer baselines. As a result, the benchmark isolates tasks that require navigating the repository structure and grounding answers in concrete code locations.
% Leveraging this benchmark, we develop a simple and practical agentic workflow for repository-level code understanding and propose a scalable training recipe that combines supervised fine-tuning (SFT) with reinforcement learning from AI feedback (RLAIF). This approach enables small open models to learn effective and economical tool use under limited context budgets. Empirically, a Qwen3-8B model trained with our recipe surpasses GPT-4o on SWE-QA-Pro by 2.3 points and substantially narrows the gap to stronger closed-source models, while producing consistently grounded, repository-aware answers.
Agentic repository-level code understanding is essential for automating complex software engineering tasks, yet the field lacks reliable benchmarks. Existing evaluations often overlook the long tail topics and rely on popular repositories where Large Language Models (LLMs) can cheat via memorized knowledge. To address this, we introduce SWE-QA-Pro, a benchmark constructed from diverse, long-tail repositories with executable environments. We enforce topical balance via issue-driven clustering to cover under-represented task types and apply a rigorous difficulty calibration process: questions solvable by direct-answer baselines are filtered out. This results in a dataset where agentic workflows significantly outperform direct answering (e.g., a ~13-point gap for Claude Sonnet 4.5), confirming the necessity of agentic codebase exploration. Furthermore, to tackle the scarcity of training data for such complex behaviors, we propose a scalable synthetic data pipeline that powers a two-stage training recipe: Supervised Fine-Tuning (SFT) followed by Reinforcement Learning from AI Feedback (RLAIF). This approach allows small open models to learn efficient tool usage and reasoning. Empirically, a Qwen3-8B model trained with our recipe surpasses GPT-4o by 2.3 points on SWE-QA-Pro  and substantially narrows the gap to state-of-the-art proprietary models, demonstrating both the validity of our evaluation and the effectiveness of our agentic training workflow.
\end{abstract}

\section{Introduction}

Repository-level code understanding is central to LLM-assisted software engineering. Real tasks require navigating many files, tracking control and data flow across modules, and verifying that implementations match intended designs. Snippet-centric QA benchmarks do not capture these behaviors, and knowledge-only prompting can hide weaknesses in navigation and grounding \citep{husain2019codesearchnet, liu2021codeqa, huang2021cosqa, lee2022cs1qa, gong2024cosqa+, sahu2024codequeries, li2024infibench}. Recent repository QA benchmarks move toward large-context, tool-using evaluation, but still focus on few projects and include many questions solvable without interacting with the codebase. \citep{abedu2025repochat, chen2025coreqa, peng2025swe,rando2025longcodebench}.

We focus on two concrete gaps. First, \textbf{limited diversity}: existing benchmarks concentrate on a few popular repositories. This leaves large parts of the natural task distribution uncovered and under-represents certain semantic categories of tasks (e.g., configuration, data plumbing, infrastructure glue shown in Appendix~\ref{sec:discussion-diversity}). Second, \textbf{uncertain need for tools}: many benchmark questions can be answered from prior knowledge or public documentation that is already covered during pretraining, so current setups do not clearly separate cases that actually requires tool-using from the cases where a single-pass, knowledge-only model with enough context and reasoning would already succeed. As a result, it is difficult to tell whether a model truly understands and operates within a particular repository, or simply recall generic knowledge.
\begin{figure*}[t]
    \vspace{-4mm}
    \centering

    \begin{subfigure}{\linewidth}
        \centering
        \includegraphics[width=\linewidth]{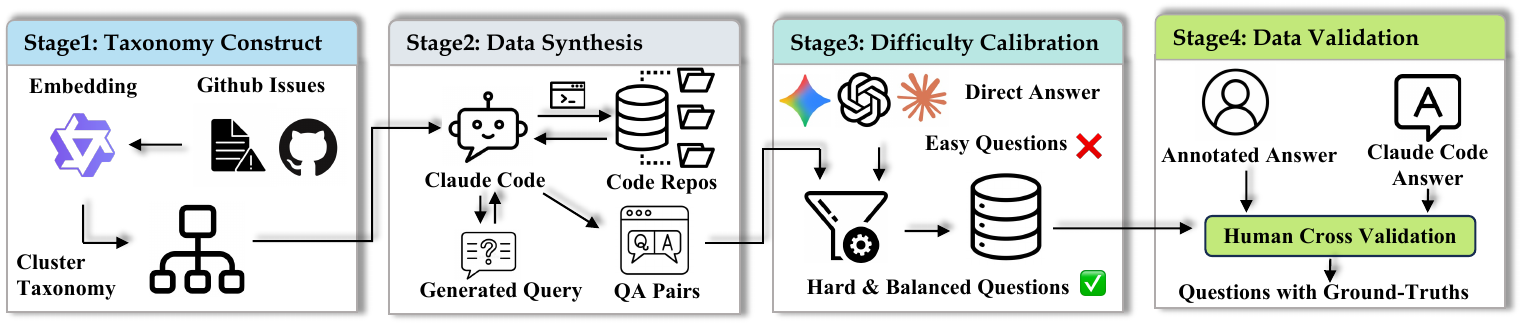}
        \caption{Benchmark Construction Pipeline}
        \label{fig:benchmark_pipeline}
    \end{subfigure}

    \vspace{0.5em}

    \begin{subfigure}{\linewidth}
        \centering
        \includegraphics[width=\linewidth]{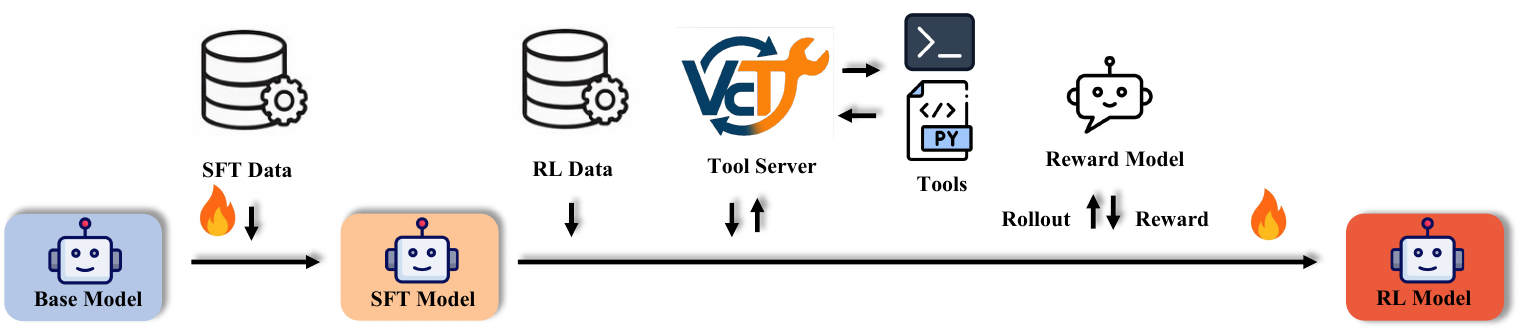}
        \caption{Training Recipe}
        \label{fig:training_recipe}
    \end{subfigure}
    \vspace{-8mm}
    \caption{SWE-QA-Pro Benchmark and Training Pipeline.}
    \label{fig:overview}
    \vspace{-4mm}
\end{figure*}

To address these issues, we introduce SWE-QA-Pro, a benchmark and training recipe for repository-level QA. On the benchmark side, we:
(i) select less-studied, long-tail repositories and ensure that each one has an executable environment so the project can be built and explored end-to-end \citep{badertdinov2025swe};
(ii) use issue texts as question seeds, embed them, and run k-means clustering to form topic groups, followed by a brief human pass to merge near-duplicates and clarify topic boundaries; and
(iii) for each topic, use a tool-using code model to propose QA items and draft answers, which are then edited by humans for correctness and repository grounding, with final benchmark QA items sampled across clusters to preserve diversity in Section ~\ref{sec:statistics}.

To reduce knowledge-only questions and make tool usage meaningful, we add a simple filtering step. For each drafted item, we compare a direct-answer baseline (no tools, single turn) with a tool-using run. If the direct-answer baseline already achieves a high score, we discard the item. This preserves questions that require locating and citing concrete code rather than recalling documentation.

On the training side, we introduce a two-stage agentic recipe for improving small open models on repository-level QA. We first apply SFT to match repository-grounded answer formats, then use RLAIF to favor answers citing concrete files and symbols \citep{lee2023rlaif}. In experiments, a tuned Qwen3-8B \citep{yang2025qwen3} trained with this SFT$\rightarrow$RLAIF recipe outperforms GPT-4o \citep{hurst2024gpt} and substantially narrows the gap to state-of-the-art proprietary models shown in Section ~\ref{sec:training} and Section ~\ref{sec:mainresults}.

In summary, we make two contributions, as illustrated in ~\autoref{fig:overview}:
\begin{itemize}[
    itemsep=0pt,
    parsep=0pt,
    topsep=0pt,
    leftmargin=1em
]
    \item \textbf{Benchmark.} We release SWE-QA-Pro, a repository-level QA benchmark built from long-tail repositories with executable environments. Questions are seeded from issues, then synthesized and grounded with a tool-using code model along with human editing, followed by filtering to remove cases solvable by strong direct-answer baselines. Compared to SWE-QA, SWE-QA-Pro covers more diverse repositories and includes more questions that truly require codebase interaction \citep{peng2025swe}.
    \item \textbf{Agent Workflow and Training Recipe.} We introduce a simple agentic workflow for repository-level QA that enables iterative codebase exploration via structured actions. Building on this workflow, we present an SFT$\rightarrow$RLAIF training recipe that significantly improves small open-source models on SWE-QA-Pro. Using this framework, Qwen3-8B surpasses GPT-4o on SWE-QA-Pro by $2.31$ points and substantially narrows the gap to several state-of-the-art proprietary models, including GPT-4.1, Claude Sonnet 4.5, and DeepSeek-V3.2 \citep{openai2025gpt41, anthropic2025claude4_5, liu2025deepseek}.
\end{itemize}

\section{SWE-QA-Pro Bench}
SWE-QA-Pro Bench is constructed through a four-stage pipeline, as illustrated in ~\autoref{fig:benchmark_pipeline}: Data Sourcing and Taxonomy, Data Synthesis and Sampling, Data Filtering and Difficulty Calibration, and Data Validation. This pipeline yields three key advantages over existing benchmarks (\autoref{tab:benchmark_comparison}):
(1) pull-request--driven clustering together with long-tail repository sampling ensures balanced coverage across diverse software engineering question types;
(2) systematic filtering against multiple strong proprietary  models removes instances solvable via memorization or pretraining artifacts, thereby isolating questions that require genuine codebase interaction;
and (3) answers cross-verified by Claude Code and human annotators provide high-quality gold ground truth, enabling reliable multi-dimensional evaluation.
% \subsection{Taxonomy Construction}

% \subsection{Problem Collection and Curation}

\begin{table*}[t]
\centering
\small
\resizebox{1\linewidth}{!}{
\begin{tabular}{lcccccc}
\toprule
\textbf{Benchmark} & \textbf{Repo-level} & \textbf{Repo Nav.} & \textbf{Multi-hop} & \textbf{Semantic Coverage} & \textbf{Diff. Calibration} & \textbf{Test Size} \\ \midrule
CodeQueries ~\citep{sahu2024codequeries} & \redmark & \redmark & \greencheck & \redmark & \redmark & 29033 \\
InfiBench ~\citep{li2024infibench} & \redmark & \redmark & \redmark & \redmark & \greencheck & 234 \\
CodeReQA ~\citep{hu2024coderepoqa}& \greencheck & \redmark & \redmark & \redmark & \redmark & 1563 \\
LongCodeQA ~\citep{rando2025longcodebench}& \greencheck & \redmark & \greencheck & \redmark & \redmark & 443 \\
SWE-QA ~\citep{peng2025swe}& \greencheck & \greencheck & \greencheck & \redmark & \redmark & 576 \\ \midrule
\textbf{SWE-QA-Pro} & \greencheck & \greencheck & \greencheck & \greencheck & \greencheck & 260 \\
\bottomrule
\end{tabular}
}
\caption{Comparison of representative code benchmarks.}
\label{tab:benchmark_comparison}  
\end{table*}

\subsection{Data Sourcing and Taxonomy}
\label{sec:taxonomy}
We conducted a large-scale analysis of the GitHub Repositories in SWE-Rebench \citep{badertdinov2025swe}. We processed $1,687,638$ issues spanning $3,468$ repositories by concatenating their titles and bodies, specifically filtering for contexts between 10 Byte and 16KB. We computed representations for these texts using Qwen3-8B-Embedding model.

To organize this data, we applied a hierarchical K-Means clustering algorithm, initializing with 10 clusters in the first layer and expanding to 50 in the second. We then utilized GPT-4.1 to extract semantic labels for each resulting cluster. These labels were refined through a human-verified taxonomy to eliminate semantic redundancy and enforce clear semantic boundaries between closely related categories, thereby reducing ambiguity and yielding 48 distinct task subclasses in Appendix \ref{sec:cluster taxonomy}. This unsupervised taxonomy serves as the foundational structure for our benchmark, ensuring it covers a wide spectrum of software engineering challenges rather than a manually cherry-picked subset.

\subsection{Data Synthesis and Sampling}
\label{subsec:synthesis}
\begin{figure*}[t]
    \centering
    % 左边的图: Issue 分布
    \begin{subfigure}[b]{0.48\textwidth}
        \centering
        \includegraphics[width=\linewidth]{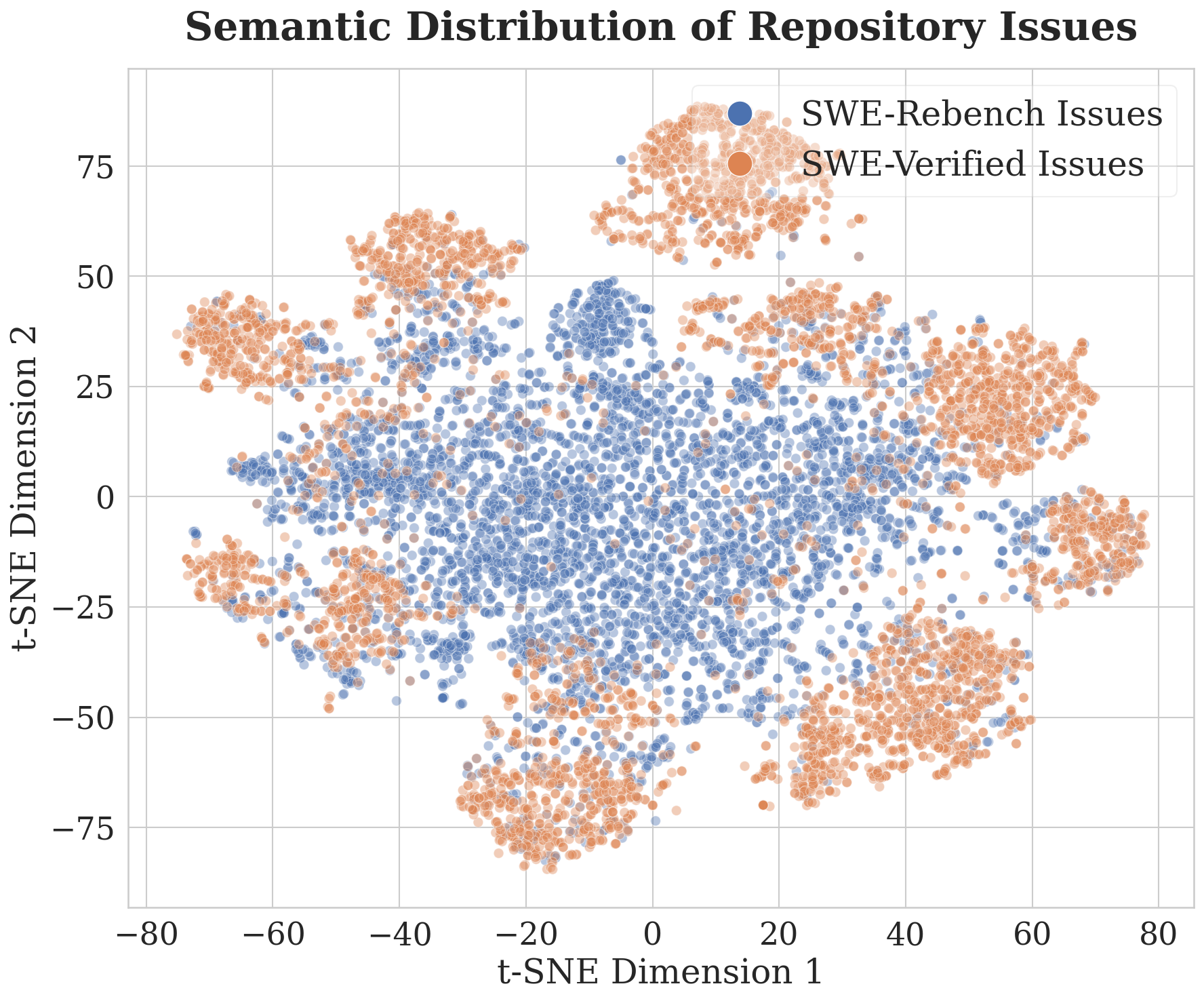}
        \caption{Semantic distribution of raw Issues: SWE-Rebench (All Repos) vs. SWE-Verified.}
        \label{fig:issue_dist}
    \end{subfigure}
    \hfill % 在两图之间通过填充空白将它们推向两端
    % 右边的图: QA Dataset 分布
    \begin{subfigure}[b]{0.48\textwidth}
        \centering
        \includegraphics[width=\linewidth]{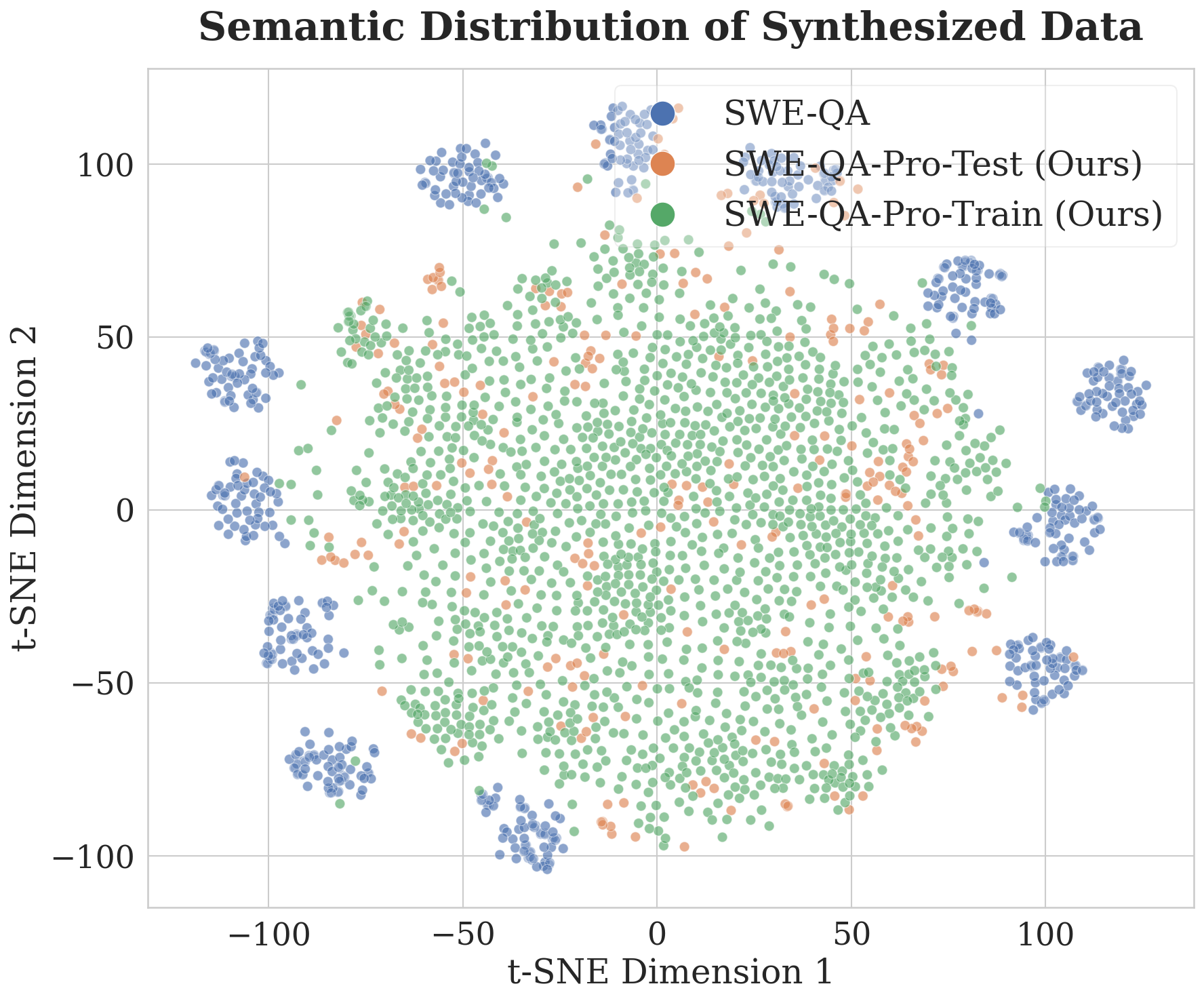}
        \caption{Semantic distribution of QA Datasets: Comparison among our Training/Test splits and SWE-QA.}
        \label{fig:qa_dist}
    \end{subfigure}
    
    \caption{t-SNE visualization of semantic distributions. \textbf{(a)} Comparison of the original issue spaces, showing the broad coverage of SWE-Rebench compared to the manually curated SWE-Verified. \textbf{(b)} The distribution of our synthesized datasets (Training and Test) demonstrates high diversity and alignment with the semantic clusters of existing benchmarks.}
    \label{fig:semantic_dist}
\end{figure*}
Leveraging the derived semantic taxonomy, we employed Claude Code to synthesize the final benchmark data. To guarantee the executability and validity of the problems, we repurposed the established sandbox environments from SWE-Rebench.

For each synthesis task, we stochastically sampled 20 existing issues from the corresponding cluster and repository to serve as reference context. The agent was then tasked with automatically exploring the codebase to generate a new, self-contained problem-solution pair aligned with the specific cluster's semantics.

We adopted different sampling strategies for the training and test sets to balance diversity with human evaluation constraints. For the Test Set, we selected 26 repositories that efficiently cover all 48 task categories, accommodating the cognitive constraints of human annotators while ensuring comprehensive evaluation. Conversely, for the Training Set, we applied uniform sampling across the entire dataset, achieving coverage of $1,484$ repositories.

~\autoref{fig:semantic_dist} illustrates the effectiveness of this pipeline: our synthesized data maintains a semantic distribution that is sufficiently diverse compared to the original repository distribution between SWE-Rebench and SWE-Verified datasets.

\subsection{Data Filtering and Difficulty Calibration}
\label{sec:filtering}
\begin{figure}[t]
    \centering
    \includegraphics[width=\linewidth]{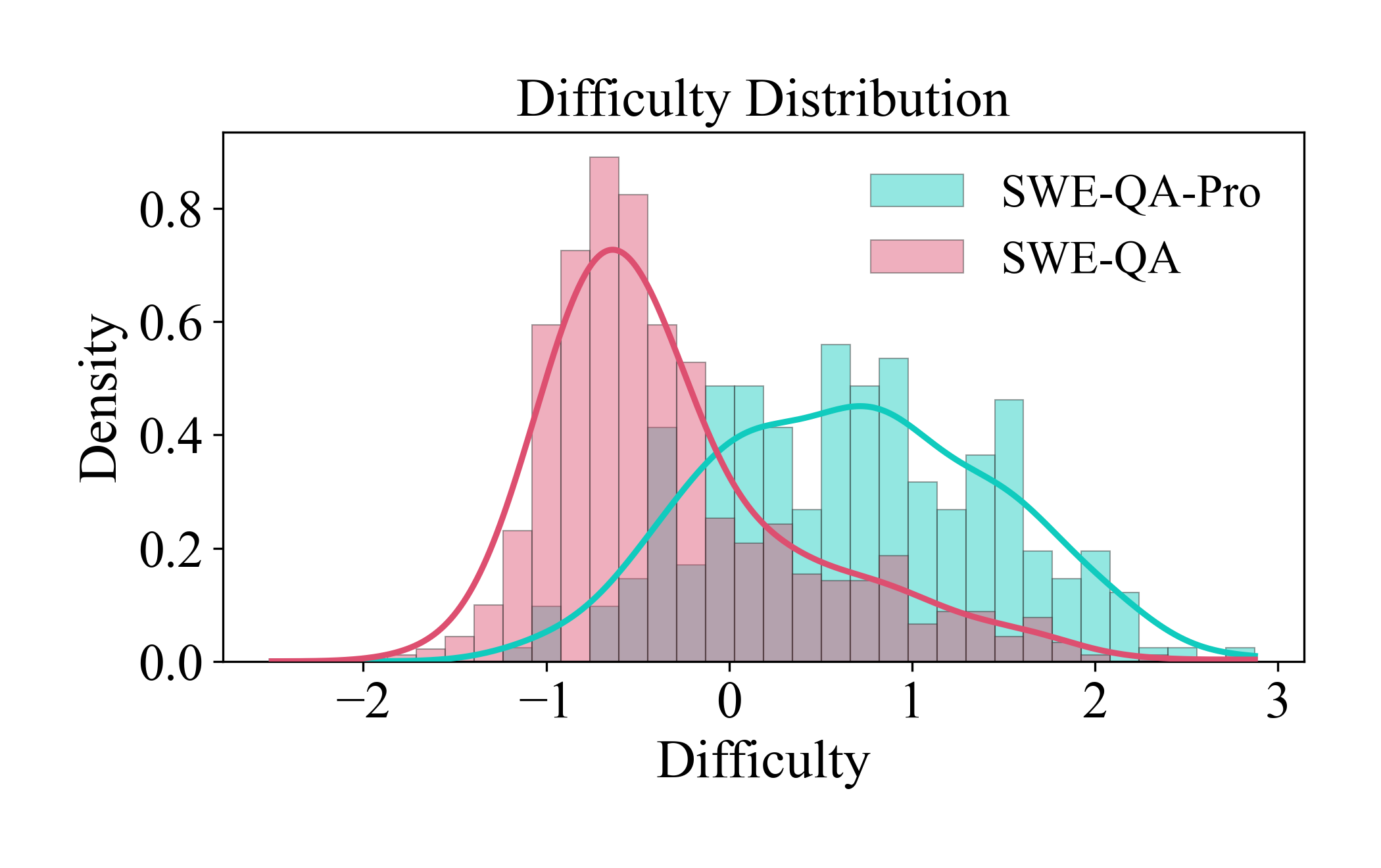}
    \vspace{-20pt}
    \caption{Difficulty Comparison between SWE-QA and SWE-QA-Pro. Higher difficulty indicates harder questions.}
    \label{fig:difficulty_kde}
    
    \vspace{-6mm}
\end{figure}
To ensure SWE-QA-Pro focuses on non-trivial, agent-essential reasoning, we apply a multi-stage filtering and calibration pipeline. We first remove multi-query prompts and perform semantic deduplication using Qwen3-8B embeddings to ensure task independence.
A key challenge in evaluating repository-level understanding is that state-of-the-art proprietary  LLMs possess extensive pretraining knowledge, enabling them to answer many software engineering questions without interacting with the codebase, reading source files, or exploring repository structure. Such questions are often associated with widely known repositories (e.g., the canonical projects in SWE-Bench \citep{jimenez2023swe}), or can be resolved by inspecting only one or a small number of files, without requiring multi-hop reasoning over the repository. Empirically, this issue is reflected in existing repo-level QA benchmarks, where the performance gap between models answering with full repository exploration and those responding without any code context is often marginal. As a result, these benchmarks may fail to accurately measure an LLM’s ability to explore codebases and perform grounded, repository-level reasoning \citep{peng2025swe}.

To mitigate the influence of memorized knowledge and filter out trivially answerable questions, we introduce a difficulty calibration procedure based on cross-model agreement. We evaluate direct (no-repository) answers produced by three strong proprietary  models, GPT-4o, Claude Sonnet 4.5, and Gemini 2.5 Pro \cite{comanici2025gemini}, and compare them against repository-grounded reference answers generated by Claude Code. Each direct answer is assessed using an LLM-as-a-Judge framework along five dimensions: correctness, completeness, relevance, clarity, and reasoning quality, as detailed in Section ~\ref{sec:setup}. For each model $m$, we aggregate scores across multiple independent runs and compute the average total score $\bar{s}_m(q)$ for question $q$.

To account for inter-model scale differences, we standardize the aggregated scores using z-score normalization:
\begin{equation}
z_m(q) = \frac{\bar{s}_m(q) - \mu_m}{\sigma_m},
\end{equation}
where $\mu_m$ and $\sigma_m$ denote the mean and standard deviation of model $m$’s scores over all questions. We then define the difficulty of a question as the negative consensus score across models:
\begin{equation}
\mathrm{Difficulty}(q) = -\frac{1}{|M|} \sum_{m \in M} z_m(q),
\end{equation}
where $M$ denotes the set of evaluated models.

Under this definition, questions that consistently receive high-quality direct answers across models are assigned low difficulty and are filtered out, while questions that remain challenging without repository interaction are retained. Using the calibrated difficulty signal, we construct the QA pairs candidates pools with cluster-level coverage and approximate balance across QA types. This calibration step ensures that SWE-QA-Pro emphasizes questions that genuinely require repository exploration and multi-step reasoning, providing a more faithful evaluation of LLM agent capabilities as shown in ~\autoref{fig:difficulty_kde}.

\subsection{Data Validation and Statistics}
\label{sec:statistics}
To ensure the quality and reliability of QA pairs and metadata, we adopt a multi-stage annotation and validation process to mitigate hallucinations and semantic ambiguity. First, \textsc{Claude Code} explores each repository in a sandbox environment to produce repository-grounded reference answers, while assigned semantic clusters and QA types are cross-checked against the taxonomy and reused for difficulty calibration.

Second, human annotators independently explore the codebase to produce answers, revise ambiguous or underspecified questions, and verify QA types and semantic clusters. Their answers are compared against the \textsc{Claude Code} references to identify missing details or errors, and only answers satisfying correctness, completeness, relevance, clarity, and reasoning quality are retained. A final expert review pass adjudicates remaining inconsistencies and further refines the answers.

The resulting SWE-QA-Pro benchmark contains 260 questions from 26 long-tail repositories, with 4–9 questions per semantic cluster and an approximately balanced distribution of QA types. Full statistics are reported in Appendix~\ref{sec:sweqapro-statistics}, with case studies in Appendix~\ref{subsec:case_compare}.

\section{SWE-QA-Pro Agent}
We introduce SWE-QA-Pro Agent, a lightweight workflow designed for repository-level code understanding in small open-source models. Unlike RAG-based approaches that require pre-built indices, our agent uses a ReAct-style loop to explore codebases directly. By combining directory traversal, keyword search, and scoped file inspection, the agent gathers evidence incrementally to reason across files under limited context budgets.
\subsection{Agent Workflow}
\label{sec:workflow}
We propose SWE-QA-Pro Agent, a ReAct-based workflow for repository-level code understanding. Prior agents such as SWE-QA-Agent primarily rely on RAG-style retrieval with limited command-line support, requiring the construction of a retrieval index while still offering insufficient capacity for genuine repository exploration. This limitation is particularly evident for open-source models, where such agents often underperform strong traditional RAG baselines with offline indexing and manually designed retrieval pipelines.

In contrast, SWE-QA-Pro Agent abandons RAG-based retrieval entirely and does not require a pre-built index. Instead, it performs direct repository exploration using explicit, length-controlled Search based on keyword matching to locate relevant files, View for scoped inspection of file contents or directory structure, and constrained read-only CommandLine actions for lightweight structural and pattern-based analysis (e.g., directory traversal, symbol matching, and line-level extraction), enabling more flexible and effective context acquisition for reasoning under limited context budgets.

The agent operates in a ReAct-style loop, where it iteratively reasons over the current context, freely selects an action, and incorporates the resulting observation until sufficient evidence is collected, at which point it terminates with Finish. Detailed algorithms are provided in the Appendix \ref{sec:algoritm}.
\subsection{Agentic Training Recipe}
\label{sec:training}
To our knowledge, existing efforts to enhance open-source LLMs for SWE-QA focus on SFT of agentic behaviors, without leveraging reinforcement learning to optimize repository-level exploration and reasoning \citep{rastogi2025devstral}. Inspired by recent advances in RL for LLMs, we propose a scalable training framework that explicitly trains agentic interaction with code repositories, leading to improved exploration and understanding.

\noindent\textbf{Training Data Construction.} Starting from the benchmark question construction pipeline, we deduplicate and obtain 1,464 raw questions, which are randomly split into 1,000 questions for SFT and 464 questions for RL. For the SFT stage, we use \texttt{Claude Sonnet 4.5} to generate 1,000 high-quality multi-turn conversation trajectories conditioned on each question and our predefined agent action space (Search, View, and read-only CommandLine tools), resulting in tool-augmented supervision data. For the RL stage, we assign each question a high-quality reference answer generated by \texttt{Claude Code}, which serves as the ground truth for reward computation.

\noindent\textbf{Two-Stage Training.} Training proceeds in two stages. In the first stage, we perform supervised fine-tuning on Qwen3-8B using 1K tool-invocation question--answer trajectories. This stage teaches the model the tool-call syntax and instills a basic understanding of tool semantics and usage patterns. In the second stage, we apply reinforcement learning to the SFT-initialized model. For each rollout, a reward model evaluates the final answer against the ground truth along five dimensions: correctness, completeness, relevance, clarity, and reasoning quality, following the same criteria used in evaluation. Since SWE-QA answers are often complex and cannot be reliably assessed by exact-match or rule-based rewards, we adopt an LLM-as-Judge reward formulation. To mitigate reward hacking, we employ a judge model distinct from the evaluation judge and assign higher weight to correctness while down-weighting clarity, discouraging fluent but incorrect answers. The final scalar reward is computed as:
\begin{equation}
\mathbf{s} = \mathrm{RM}(\hat{a}, a^{*}) \in [1,10]^5,
\end{equation}
\begin{equation}
r = \frac{\mathbf{w}^{\top}\mathbf{s}}{10},
\quad
\mathbf{w} = (0.3, 0.2, 0.2, 0.1, 0.2).
\end{equation}

\begin{table*}[ht]
\centering
\small
\begin{tabular}{lcccccc}
\toprule
\multirow{2}[2]{*}{\textbf{Model}} & \multicolumn{5}{c}{\textbf{Evaluation Metrics}} & \multirow{2}[2]{*}{\hspace{2mm} \textbf{Overall}} \\ \cmidrule{2-6} 
& \textbf{Correctness} & \textbf{Completeness} & \textbf{Relevance} & \textbf{Clarity} & \textbf{Reasoning} & \\ 
\midrule
\multicolumn{7}{c}{\textit{Proprietary LLMs}} \\
\midrule
Gemini 2.5 Pro & 2.51 & 2.13 & 8.66 & 8.02 & 4.16 & 25.48 \\
Gemini 2.5 Pro + Agent & \uline{7.12} & 6.25 & 8.91 & \textbf{9.34} & \uline{7.84} & \uline{39.46} \\
GPT-4.1 & 3.42 & 2.38 & \textbf{9.02} & \uline{9.23} & 4.68 & 28.74 \\
GPT-4.1 + Agent & 6.86 & 5.90 & 8.89 & 9.13 & 7.68 & 38.47 \\
GPT-4o & 3.08 & 2.11 & \uline{8.96} & 8.79 & 3.64 & 26.58 \\
GPT-4o + Agent & 5.59 & 4.49 & 8.55 & 8.16 & 6.29 & 33.08 \\
DeepSeek V3.2 & 3.19 & 2.32 & 8.83 & 8.83 & 4.39 & 27.55 \\
DeepSeek V3.2 + Agent & 6.94 & \uline{6.49} & 8.78 & 8.72 & 7.76 & 38.69 \\
Claude Sonnet 4.5 & 3.34 & 2.74 & 8.65 & 8.12 & 4.84 & 27.69 \\
Claude Sonnet 4.5 + Agent & \textbf{7.34} & \textbf{7.36} & 8.88 & 9.03 & \textbf{8.06} & \textbf{40.67} \\
\midrule
\multicolumn{7}{c}{\textit{Open-Source LLMs}} \\
\midrule
Qwen3-8B & 2.84 & 2.16 & 8.59 & 8.66 & 4.36 & 26.61 \\
Qwen3-8B + Agent & 4.52 & 3.77 & 8.29 & 7.83 & 5.62 & 30.03 \\
Qwen3-32B & 3.04 & 2.41 & \uline{8.71} & \uline{8.74} & 5.02 & 27.91 \\
Qwen3-32B + Agent & \uline{4.99} & \uline{4.21} & 8.50 & 8.16 & \uline{6.22} & \uline{32.08} \\
Llama-3.3-70B-Instruct & 2.34 & 1.75 & 8.68 & 8.47 & 3.08 & 24.32 \\
Llama-3.3-70B-Instruct + Agent & 2.84 & 2.11 & 8.09 & 7.18 & 3.51 & 23.73 \\
Devstral-Small-2-24B-Instruct & 2.65 & 2.14 & 8.57 & 8.31 & 3.77 & 25.44 \\
Devstral-Small-2-24B-Instruct + Agent & \textbf{6.61} & \textbf{5.66} & \textbf{8.81} & \textbf{9.09} & \textbf{7.13} & \textbf{37.30} \\
\midrule
\multicolumn{7}{c}{\textit{Finetuned LLMs}} \\
\midrule
SWE-QA-Pro-8B (SFT) & 2.56 & 2.01 & 8.37 & 7.93 & 3.55 & 24.42 \\
SWE-QA-Pro-8B (SFT) + Agent & \uline{5.66} & \uline{5.45} & \uline{8.40} & \uline{8.21} & \uline{6.61} & \uline{34.34} \\
SWE-QA-Pro-8B (SFT+RL) & 2.54 & 2.04 & 8.28 & 7.92 & 3.55 & 24.34 \\
SWE-QA-Pro-8B (SFT+RL) + Agent & \textbf{5.96} & \textbf{5.66} & \textbf{8.51} & \textbf{8.44} & \textbf{6.83} & \textbf{35.39} \\
\bottomrule
\end{tabular}
\caption{SWE-QA-Pro Bench evaluation results. ``+agent'' denotes models using the SWE-QA-Pro agent framework. Best results per scale are shown in \textbf{bold}, with \underline{second-best} underlined.}
\vspace{-4mm}
\label{tab:main-results}
\end{table*}

where $\hat{a}$ is the generated answer, $a^{*}$ the ground-truth reference, and $\mathbf{s}$ denotes scores for five evaluation dimensions. We optimize the policy using the GRPO algorithm \citep{shao2024deepseekmath}, where rewards are normalized within each rollout group before computing policy gradients. This stage encourages the model to converge toward rollouts that produce high-quality, fact-grounded final answers.

\section{Experiments}

\subsection{Experimental Setup}
\label{sec:setup}
\noindent\textbf{Model selection} We evaluate 11 LLMs, including proprietary models, GPT-4o, GPT-4.1, Claude Sonnet 4.5, Gemini 2.5 Pro, DeepSeek-V3.2, open-source models, Qwen3-8B/32B, Devstral-Small-2-24B-Instruct \citep{rastogi2025devstral}, LLaMA-3.3-70B-Instruct \citep{dubey2024llama}, and two variants of SWE-QA-Pro 8B trained with SFT and SFT+RL. All models are evaluated under both direct answering and agent-based reasoning using the SWE-QA-Pro Agent workflow.

\noindent\textbf{Inference and Training Setup} All inference uses temperature 0, a maximum of 25 turns, and a 32k context window on NVIDIA A100 80GB GPUs. SFT and RL are implemented using SWIFT \citep{zhao2025swift} and Verl-Tool \citep{jiang2025verltool}, respectively. Hyperparameters are provided in Appendix \ref{sec:hyperparameters}.

\noindent\textbf{Evaluation Metrics} We follow the LLM-as-Judge protocol of SWE-QA, including strict judge–candidate separation, anonymization, and randomized answer order. Compared to SWE-QA, we require explicit file-path and line-number references and use a stricter judge prompt to enable finer-grained score differentiation. Each answer is scored independently three times, and scores are averaged to reduce variance. Full prompts are provided in Appendix \ref{sec:prompts}.

\subsection{Main Results}
\label{sec:mainresults}
~\autoref{tab:main-results} summarizes the evaluation results across all LLMs. As shown in ~\autoref{tab:main-results}, there is a substantial performance gap between the direct-answer setting and the agent-based workflow, particularly on correctness, completeness, and reasoning quality. This gap highlights the effectiveness of our difficulty calibration and underscores the critical role of the SWE-QA-Pro agent in enabling high-quality, repository-grounded reasoning.

\begin{figure*}[ht]
  \centering
  \vspace{-4mm}
  \includegraphics[width=\textwidth]{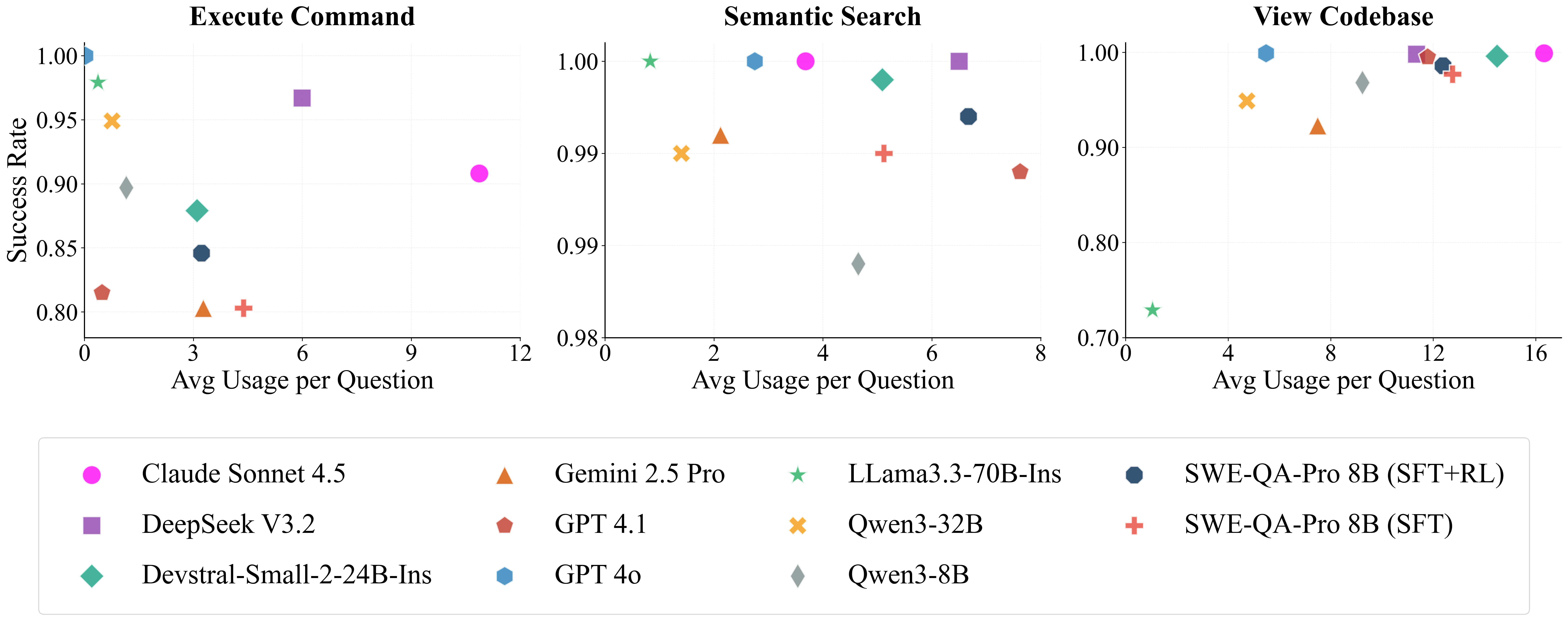}
  \vspace{-3mm}
  \caption{
  Tool usage behavior across models on SWE-QA-Pro.
  }
  \label{fig:tool-usage}
  \vspace{-6mm}
\end{figure*}

\noindent\textbf{Overall Performance.} Among all evaluated models, Claude Sonnet 4.5 achieves the highest overall score, reflecting strong repository-level code understanding and effective tool use.
Despite its smaller scale and limited training data, SWE-QA-Pro 8B outperforms many open-source baselines as well as GPT-4o with case study in Appendix \ref{subsec:case_session}, and performs competitively with larger agentic models such as Devstral Small 2 24B Instruct.
These results highlight that explicitly training agentic capabilities for repository-level QA can be more impactful than scaling model size alone.

\noindent\textbf{Breakdown Results Analysis.} Appendix~\ref{sec:breakdown} details model performance across repositories, semantic clusters, and question types. First, analyzing question types (\autoref{tab:rq3_main},~\autoref{tab:rq3_main_qwen}) reveals that localization-oriented questions yield consistently high scores with low variance, as they focus on identifying specific files, identifiers, or execution points. Conversely, causal and explanatory questions are significantly more challenging, particularly those involving design rationale, trade-offs, or implicit dependencies, which require multi-file evidence integration and global semantic reasoning. Procedural questions fall in between: concrete implementation tasks are tractable, whereas system-level inquiries remain difficult due to their reliance on holistic understanding.

Regarding repositories and semantic clusters in Appendix \ref{subsec:breakdown_reponame} and \ref{subsec:breakdown_cluster}, configuration and workflow management areas involving dependency injection, CLI argument handling, and packaging prove consistently difficult. These clusters characterize configuration-driven repositories like jsonargparse, checkov, and yt-dlp, where answering requires reasoning over implicit control flow, cross-file propagation, and runtime behavior not localized to single files. In contrast, clusters with explicitly encoded, localized logic, such as Unicode/data parsing, protocol/API compatibility, filesystem config, and visualization, are easier. Consequently, repositories concentrating on these areas (docker-py, pint, mkdocs, seaborn) achieve higher, stable performance, benefiting from structurally explicit and locally grounded code.

\subsection{Tool Usage Analysis}

~\autoref{fig:tool-usage} correlates tool proficiency with repository-level QA performance. Models with lower scores, such as LLaMA-3.3-70B-Instruct and GPT-4o, suffer from weak tool usage that limits context retrieval and global understanding. Conversely, Claude Sonnet 4.5 excels by leveraging the highest volume of tool calls, translating robust exploration into superior answer quality. Gemini 2.5 Pro, however, remains competitive with fewer calls, indicating that internal reasoning enables efficient, selective tool use; this underscores that reasoning is vital alongside tool capacity. Additionally, post-RL improvements in SWE-QA-Pro 8B demonstrate that RL fosters effective, judicious execution rather than merely inflating tool-call frequency.

\subsection{Training Strategy Analysis}
\begin{figure}[ht]
    \centering
    \includegraphics[width=\linewidth]{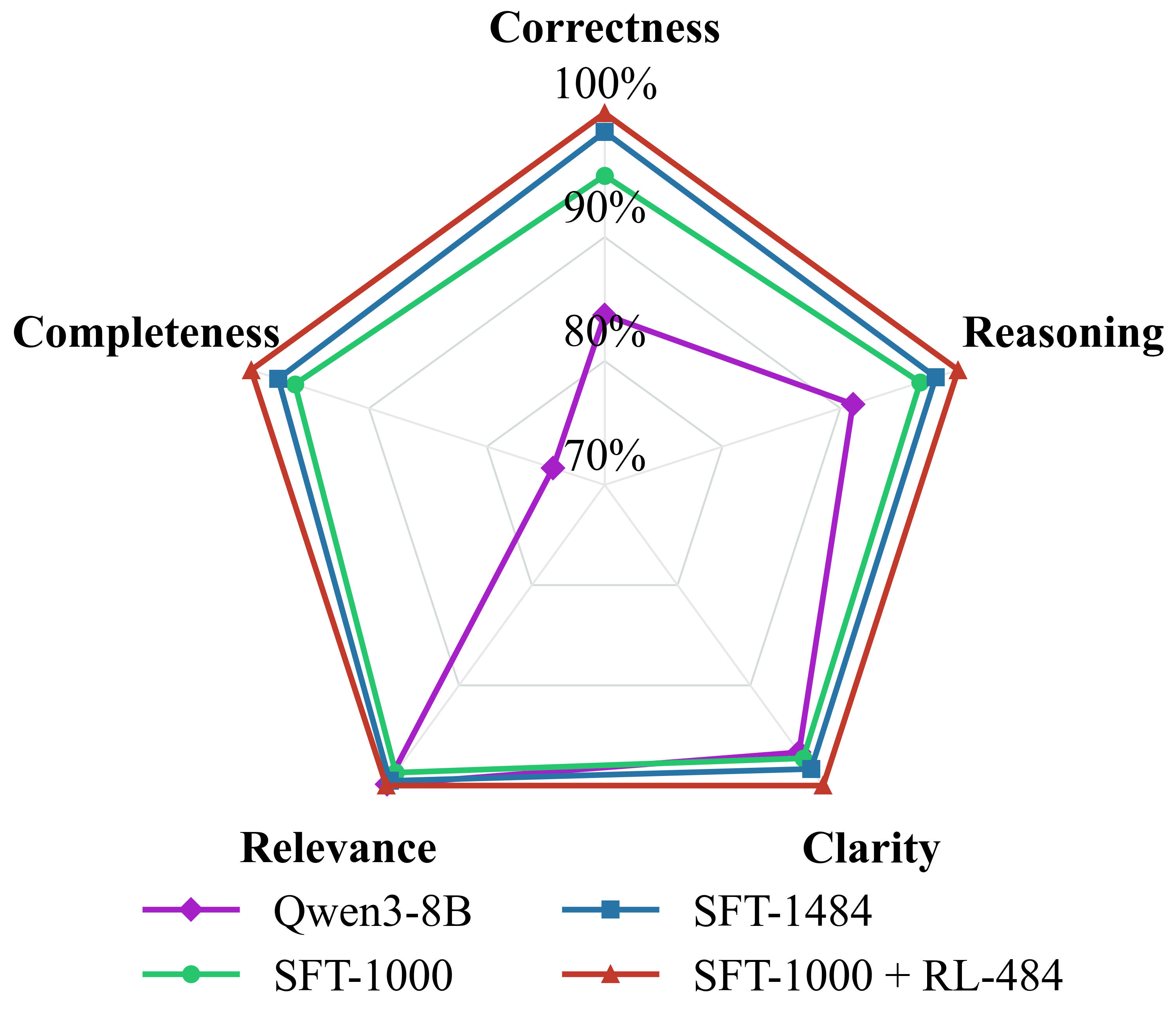}
    \vspace{-4mm}
    \caption{Effect of Training Strategy. We compare SFT at different scales and a two-stage SFT→RLAIF setting.}
    \label{fig:train_compare}
    \vspace{-8mm}
\end{figure}

~\autoref{fig:train_compare} compares different training strategies under the same model backbone. SFT-1000 and SFT-1464 denote supervised fine-tuning on 1,000 and 1,464 tool-call trajectories, respectively, while SFT-1000 + RL-464 represents a two-stage setting that initializes the model with 1,000 SFT trajectories and then applies reinforcement learning on an additional 464 QA pairs. Increasing the amount of supervised fine-tuning data from 1,000 to 1,464 trajectories yields consistent but modest improvements across most evaluation dimensions. In contrast, introducing reinforcement learning after SFT leads to a more pronounced performance gain. In particular, SFT-1000 + RL-464 achieves substantially higher scores in both Correctness and Completeness than SFT-only variants, including SFT-1464. This indicates that reinforcement learning does not simply replicate the effect of scaling supervised data, but instead introduces a qualitatively different optimization signal that further unlocks the potential of the SFT-initialized model, encouraging more accurate and more comprehensive answers. Overall, these results suggest that RL provides complementary supervision beyond SFT, especially effective at refining factual precision and answer coverage.

\section{Related Work}
\noindent\textbf{Code-centric and Repository-level QA Benchmarks.}
Existing code and repository QA benchmarks focus on localized or context-limited settings, where questions can be answered from snippets, APIs, or documentation.
Representative datasets such as CodeQueries, CS1QA, CoSQA, CoSQA+, and CodeSearchNet emphasize element-level reasoning or retrieval over individual functions, deliberately avoiding cross-file dependencies and repository structure \citep{sahu2024codequeries, lee2022cs1qa, huang2021cosqa, gong2024cosqa+, husain2019codesearchnet}. InfiBench extends evaluation to free-form coding-related questions across languages, but remains knowledge- and snippet-centric rather than repository-grounded \citep{li2024infibench}.

More recent efforts move toward repository-scale evaluation. LongCodeBench relies on long context windows to ingest large codebases \citep{rando2025longcodebench}, while RepoChat uses offline indexing for structured retrieval \citep{abedu2025repochat}. SWE-QA formulates repository understanding as a QA task \citep{peng2025swe}, but does not explicitly separate cases solvable by standard retrieval. In contrast, SWE-QA-Pro targets long-tail, executable repositories and filters out retrieval-solvable queries, isolating scenarios that require interactive code exploration.

% More recent efforts move toward repository-aware evaluation.
% LongCodeBench stresses extremely long context windows by exposing models to large portions of codebases, prioritizing context scale over interactive exploration \citep{rando2025longcodebench}, while systems such as RepoChat rely on offline indexing and structured representations to answer repository-related questions \citep{abedu2025repochat}.
% Most closely related, SWE-QA explicitly frames repository-level code understanding as a QA task \citep{peng2025swe}.
% In contrast, SWE-QA-Pro focuses on long-tail, executable repositories and filters out questions solvable by direct-answer baselines, isolating cases that genuinely require interaction with the codebase.

\noindent\textbf{Repository-level Agents.} Agents in software engineering largely target generative tasks, including issue resolution, program repair, and code generation \citep{jimenez2023swe, yang2024swe, zhang2024autocoderover, da2025agent, li2025swe, bi2024iterative, bairi2024codeplan}. In these domains, exploration is implicitly shaped by generation objectives rather than comprehension. Conversely, repository-level QA demands strict code navigation and understanding. Prior approaches, such as SWE-QA-Agent, rely on inference-time heuristics for tool use, often underperform retrieval-augmented generation (RAG) baselines due to unoptimized navigation \citep{peng2025swe}. We address this limitation by explicitly training repository exploration policies, bridging the gap between passive retrieval and active agentic navigation.

% \noindent\textbf{Repository-level Agents.}
% Repository-level agents have been studied across multiple software engineering tasks, including issue resolution and program repair \citep{jimenez2023swe, yang2024swe, zhang2024autocoderover, da2025agent, li2025swe}, code and issue localization \citep{jiang2025cosil, chen2025locagent, liu2025graphlocator}, repository-level code completion \citep{bi2024iterative, bairi2024codeplan}, and unit test generation \citep{cheng2025agentic, ahmed2025execution}.
% These settings predominantly emphasize generation or repair outcomes, where exploration policies are often shaped indirectly through task-level objectives. By contrast, repository-level question answering isolates code understanding and navigation from code generation.
% In this setting, existing approaches such as SWE-QA-Agent rely primarily on inference-time tool invocation without explicitly training repository exploration policies, leading to weak performance on open-source models compared to traditional RAG baselines \citep{peng2025swe}. SWE-QA-Pro addresses this gap by explicitly training repository exploration behaviors and evaluating them under realistic, constrained-context settings.

\section{Conclusion}
% We study repository-level QA in the context of real-world software engineering, where answering a question often requires navigating multiple files and grounding claims in concrete implementations. However, many existing benchmark questions can still be answered without truly interacting with the codebase and provide limited semantic coverage. To clarify this boundary, we introduce SWE-QA-Pro, which expands coverage to long-tail, executable repositories and explicitly filters out items that strong direct-answer baselines can already solve. This design better isolates cases where repository navigation and code grounding are necessary.

% Paired with SWE-QA-Pro, we present a simple SFT$\rightarrow$RLAIF training recipe that improves the ability of small open models to produce repository-grounded answers. While the benchmark remains limited in scale and focuses on read-only interaction, we hope SWE-QA-Pro provides a cleaner testbed for studying when and how codebase interaction truly matters in repository-level QA.

In this work, we address the challenge of evaluating and training Large Language Models for repository-level code understanding, where reliance on memorized knowledge often masks deficits in genuine exploration capabilities. By introducing SWE-QA-Pro, we establish a rigorous testbed that enforces semantic diversity through long-tail repositories and systematically filters out questions solvable by direct answering. The substantial performance gap observed between direct and agentic baselines on this benchmark confirms that our design successfully isolates tasks requiring authentic codebase navigation and evidence grounding.

Beyond evaluation, we show that agentic capabilities can be learned using a scalable framework. We propose a two-stage SFT$\rightarrow$RLAIF training recipe enabled by a synthetic data pipeline. A Qwen3-8B model trained with this recipe surpasses GPT-4o on SWE-QA-Pro and substantially narrows the gap to state-of-the-art proprietary models. 
We hope SWE-QA-Pro catalyzes future research toward active, grounded repository reasoning.

\section*{Limitations}
% The objective of SWE-QA-Pro is to provide a challenging and realistic benchmark, together with a practical agentic training recipe, for repository-level code understanding. While our design emphasizes semantic diversity, executable environments, careful human verification, and explicit filtering of questions solvable by pretraining knowledge alone, the current benchmark remains limited in scale, with 260 questions from 26 repositories, reflecting annotation and validation costs rather than methodological constraints. On the training side, although our experiments use a modest amount of high-quality supervision, the proposed SFT$\rightarrow$RLAIF framework is inherently scalable, as both data synthesis and reward-based optimization naturally extend to larger repository collections. We leave a systematic empirical study of scaling behavior with increased data and compute to future work.
The objective of SWE-QA-Pro is to provide a challenging benchmark and practical training recipe, yet we identify three limitations. First, despite employing semantic embedding clustering to maximize topical coverage, the benchmark is constrained to 260 questions from 26 repositories due to the high cost of expert human verification; this scale may not fully capture the extreme long-tail diversity of the software ecosystem, and the embedding models used for clustering could inadvertently introduce latent biases into the taxonomy. Second, the benchmark is currently restricted to the Python ecosystem due to the strict requirement for executable sandboxes to verify agent actions, though our data synthesis and agentic training pipeline is inherently language-agnostic and can be readily extended to other languages with compatible runtime environments. Third, our RLAIF training objectives share a similar distribution with evaluation metrics as both rely on LLM-as-a-Judge frameworks, creating a potential risk of reward hacking where models optimize for judge preferences rather than objective correctness. While preliminary case studies did not reveal significant gaming behaviors, this proximity suggests that future research should prioritize investigating more robust training methodologies—such as process supervision or diverse reward modeling—to mitigate such alignment risks.

\bibliography{custom}

\appendix
\appendix
\newpage
\onecolumn
\clearpage
\phantomsection
\label{list:list_of_appendix}
\DoToC

\clearpage
\clearpage
\section{Cluster Coverage of SWE-QA Bench}
\label{sec:discussion-diversity}

\begin{table}[H]
\centering
\small
\label{tab:cluster-coverage-01}
\resizebox{\linewidth}{!}{
\begin{tabular}{l >{\arraybackslash}m{3.8cm} c c c c c}
\toprule
Cluster ID & Cluster Name & Original Ratio (\%) & Test Ratio (\%) & Coverage ($\times$) & Origin Count & Test Count \\
\midrule
0.0 & Unicode / formatting / parsing / data validation & 2.04 & 0.00 & 0.00 & 17,185 & \textbf{0} \\
0.1 & SQL / structured grammar / templating Engine & 2.67 & 1.89 & 0.71 & 22,489 & 10 \\
0.2 & CLI args / syntax / regex / command completion & 2.08 & 3.22 & 1.55 & 17,518 & 17 \\
0.3 & file import/export / metadata / sorting / recurrence & 1.89 & 0.19 & 0.10 & 15,940 & \textbf{1} \\
\midrule
1.0 & data formats / FITS / Astropy / units / WCS & 0.93 & 8.14 & 8.78 & 7,825 & 43 \\
1.1 & pandas / parquet / datetime / Dask / table schema & 1.77 & 0.57 & 0.32 & 14,917 & \textbf{3} \\
1.2 & numpy / dask arrays / dtype / array serialization / parallel & 1.96 & 8.33 & 4.24 & 16,571 & 44 \\
1.3 & coordinate transform / image processing / IO / numeric precision & 3.99 & 0.00 & 0.00 & 33,660 & \textbf{0} \\
\midrule
2.0 & protocol serialization / encoding / headers / API compatibility & 1.25 & 0.00 & 0.00 & 10,573 & \textbf{0} \\
2.1 & dependency injection / config / attrs / API design & 2.27 & 7.20 & 3.17 & 19,168 & 38 \\
2.2 & dataclass / schema validation / enums / OpenAPI & 1.68 & 0.76 & 0.45 & 14,190 & 4 \\
2.3 & type and attribute errors / initialization / CLI workflow & 2.00 & 0.19 & 0.09 & 16,837 & \textbf{1} \\
2.4 & type hints / mypy / typing system / code generation & 1.71 & 1.52 & 0.89 & 14,436 & 8 \\
\midrule
3.0 & Python version / imports / deprecation / conflicts & 1.93 & 0.00 & 0.00 & 16,300 & \textbf{0} \\
3.1 & install / virtual env / OS / hardware requirements / cloud deploy & 2.14 & 0.00 & 0.00 & 18,035 & \textbf{0} \\
3.2 & artifacts / distribution format / repository management / post-install state & 2.09 & 0.57 & 0.27 & 17,677 & 3 \\
3.3 & extensibility / configuration framework / plugin architecture & 2.96 & 0.00 & 0.00 & 24,998 & \textbf{0} \\
\midrule
4.0 & version control / Docker / build cache & 1.81 & 0.00 & 0.00 & 15,274 & \textbf{0} \\
4.1 & release management / changelog / license / community & 1.98 & 0.00 & 0.00 & 16,692 & \textbf{0} \\
4.2 & documentation / MkDocs / user tutorials & 3.11 & 5.30 & 1.70 & 26,274 & 28 \\
4.3 & async refactor / migration / logging infra / i18n & 2.37 & 0.38 & 0.16 & 19,972 & \textbf{2} \\
4.4 & CI pipelines / coverage / lint / GitHub Actions / security checks & 1.30 & 0.19 & 0.15 & 11,011 & \textbf{1} \\
\bottomrule
\end{tabular}}
\caption{Cluster Coverage (1.0 - 4.4): SWE-Rebench (Original) vs. SWE-QA (Test). \newline \centering \hyperref[list:list_of_appendix]{[Back to Appendix Contents]}}
\end{table}

\newpage
\setlength{\tabcolsep}{3pt}
\renewcommand{\arraystretch}{0.7}
\begin{table}[H]
\centering
\small
\label{tab:cluster-coverage-02}
\resizebox{\linewidth}{!}{
\begin{tabular}{l >{\arraybackslash}m{3.8cm} c c c c c}
\toprule
Cluster ID & Cluster Name & Original Ratio (\%) & Test Ratio (\%) & Coverage ($\times$) & Origin Count & Test Count \\
\midrule
5.0 & asyncio / async context / resource cleanup & 2.11 & 0.19 & 0.09 & 17,811 & \textbf{1} \\
5.1 & multiprocessing / advanced runtime / concurrency / heterogeneous compute & 1.07 & 0.00 & 0.00 & 9,034 & \textbf{0} \\
5.2 & runtime error handling / DB transactions / retry / logging system & 2.48 & 0.76 & 0.31 & 20,891 & 4 \\
5.3 & threading / execution limits / scheduling / memory / timeout & 2.17 & 0.76 & 0.35 & 18,278 & 4 \\
5.4 & connection lifecycle / protocol handling / low-level failures & 2.08 & 0.19 & 0.09 & 17,569 & \textbf{1} \\
5.5 & parallel execution / distributed frameworks / task graphs & 1.72 & 0.00 & 0.00 & 14,511 & \textbf{0} \\
\midrule
6.0 & file paths / filesystem permissions / symlinks / env config / cache system & 2.13 & 0.76 & 0.36 & 17,982 & 4 \\
6.1 & unit testing / mocking / test automation & 2.05 & 8.90 & 4.34 & 17,323 & 47 \\
6.2 & build pipeline / doc building / Sphinx / cloud provisioning & 3.30 & 0.00 & 0.00 & 27,858 & \textbf{0} \\
6.3 & compiler toolchain / cross-compile / env vars / code quality analysis & 1.42 & 0.00 & 0.00 & 11,984 & \textbf{0} \\
\midrule
7.0 & API integration / sync / performance / DB / SDK & 2.62 & 0.95 & 0.36 & 22,067 & 5 \\
7.1 & media download / playlist / metadata / client-side proxy config & 1.13 & 0.00 & 0.00 & 9,532 & \textbf{0} \\
7.2 & auth systems / deployment / extension plugins / cloud services & 2.21 & 2.46 & 1.11 & 18,668 & 13 \\
7.3 & AWS / Azure / K8s / container security / IAM policy & 2.20 & 0.00 & 0.00 & 18,531 & \textbf{0} \\
7.4 & reverse proxy / URL routing / websocket / CDN / streaming & 1.42 & 15.34 & 10.79 & 12,000 & 81 \\
7.5 & OAuth / JWT / SSL / access control / user sessions/ token lifecycle & 4.00 & 1.33 & 0.33 & 33,762 & 7 \\
\midrule
8.0 & tensors / training / GPU / ML experiment logging / tuning & 2.20 & 2.08 & 0.95 & 18,525 & 11 \\
8.1 & ML analytical visualization / Fourier / ML animation / calibration & 2.76 & 1.33 & 0.48 & 23,278 & 7 \\
8.2 & time series / feature engineering / explainability methods / behavioral analysis / computational semantics & 0.90 & 2.46 & 2.73 & 7,606 & 13 \\
8.3 & data parallel / compression / ML plugin / indexing & 2.39 & 2.65 & 1.11 & 20,182 & 14 \\
8.4 & bayesian models / MCMC / statistics / reproducibility & 1.67 & 1.14 & 0.68 & 14,066 & 6 \\
8.5 & ML APIs / decorators / metrics / optimization strategies & 2.15 & 10.61 & 4.94 & 18,116 & 56 \\
\midrule
9.0 & UI layout / CSS / markdown / table extraction / frontend security & 2.30 & 0.38 & 0.16 & 19,450 & 2 \\
9.1 & plotting systems / widgets / maps / UI animation / usability & 1.78 & 4.92 & 2.77 & 14,995 & 26 \\
9.2 & runtime UI config / UI permission management / upload handling / customization / user-facing runtime extensibility & 2.09 & 0.19 & 0.09 & 17,643 & \textbf{1} \\
9.3 & 3D rendering / legends / color mapping / visualization formatting & 1.73 & 4.17 & 2.41 & 14,615 & 22 \\
\bottomrule
\end{tabular}
}
\caption{Cluster Coverage (5.0 - 9.3): SWE-Rebench (Original) vs. SWE-QA (Test). \newline \centering \hyperref[list:list_of_appendix]{[Back to Appendix Contents]}}
\end{table}
\renewcommand{\arraystretch}{1.0}
\setlength{\tabcolsep}{6pt}

\clearpage
\section{SWE-QA-Pro Agent Algorithm}
\label{sec:algoritm}

\begin{algorithm}[H]
\caption{SWE-QA-Pro Agent}
\label{alg:sweqa-pro-agent}
\begin{algorithmic}[1]
\Require User query $Q$, repository $R$
\Ensure Final answer $A$

\State \textbf{/* Phase 1: Initialization */}
\State $context \gets [\ ]$
\State $thought \gets Analyze(Q)$
\State $context \gets context \cup \{SemanticSearch(Q, R)\}$

\Statex
\State \textbf{/* Phase 2: Iterative ReAct Loop */}
\State $max\_iterations \gets N$
\For{$i \gets 1$ \textbf{to} $max\_iterations$}
    \State $thought \gets Reason(context, Q)$
    \State $action \gets SelectAction(thought)$

    \If{$action = SemanticSearch$}
        \State $output \gets Execute(SemanticSearch)$
    \ElsIf{$action = ViewCodebase$}
        \State $output \gets Execute(ViewCodebase)$
    \ElsIf{$action = ExecuteCommand$}
        \State $output \gets Execute(ExecuteCommand)$
    \EndIf

    \State $context \gets context \cup \{output\}$

    \If{$SufficientEvidence(context, Q)$ \textbf{or} $i = max\_iterations$}
        \State \textbf{break}
    \EndIf
\EndFor

\Statex
\State \textbf{/* Phase 3: Finalization */}
\State $A \gets Synthesize(context, Q)$
\State \Return $A$
\end{algorithmic}
\end{algorithm}

\clearpage
\section{Training Hyperparameters}
\label{sec:hyperparameters}

\begin{table}[H]
\centering
\resizebox{0.44\linewidth}{!}{
\begin{tabular}{p{0.33\textwidth} l}
\toprule
\textbf{Hyperparameter} & \textbf{Value} \\
\midrule
Precision & bfloat16 \\
Max sequence length & 32,768 \\
Optimizer & AdamW \\
Learning rate & $5 \times 10^{-6}$ \\
Weight decay & 0.05 \\
LR scheduler & Cosine \\
Warmup ratio & 0.05 \\
Batch size (per device) & 1 \\
Gradient accumulation & 2 \\
Epochs & 4 \\
Agent template & Hermes \\
\bottomrule
\end{tabular}
}
\caption{Hyperparameters for SFT of SWE-QA-Pro 8B. \centering \hyperref[list:list_of_appendix]{[Back to Appendix Contents]}}
\label{tab:sft-hparams}
\end{table}

\begin{table}[H]
\centering
\resizebox{0.46\linewidth}{!}{
\begin{tabular}{p{0.33\textwidth} l}
\toprule
\textbf{Hyperparameter} & \textbf{Value} \\
\midrule
Max turns & 25 \\
Max prompt length & 2,048 \\
Max response length & 8,192 \\
Max observation length & 28,000 \\
Temperature & 1.0 \\
Top-p & 1.0 \\
Number of rollouts (n) & 8 \\
KL loss coefficient & 0.02 \\
KL loss type & Low-variance KL \\
Entropy coefficient & 0 \\
Actor learning rate & $1 \times 10^{-6}$ \\
Batch size & 8 \\
PPO mini-batch size & 8 \\
Strategy & FSDP \\
Max model length & 32,768 \\
\bottomrule
\end{tabular}
}
\caption{Hyperparameters for RL of SWE-QA-Pro 8B. \centering \hyperref[list:list_of_appendix]{[Back to Appendix Contents]}}
\label{tab:rl-hparams}
\end{table}

\clearpage
\section{Prompts}
\label{sec:prompts}

\begin{promptbox}[Prompt Template for LLM-as-Judge]{prompt_purple}
\textbf{Model: GPT-5}\\ \\
You are a professional evaluator. Please rate the candidate answer against the reference answer based on five criteria.
Evaluation Criteria and Scoring Guidelines (each scored 1 to 10):\\

1. Correctness:\\
10 — Completely correct; core points and details are accurate with no ambiguity.\\
8-9 — Mostly correct; only minor details are slightly inaccurate or loosely expressed.\\
6-7 — Partially correct; some errors or omissions, but main points are generally accurate.\\
4-5 — Several errors or ambiguities that affect understanding of the core information.\\
2-3 — Many errors; misleading or fails to convey key information.\\  
1 — Serious errors; completely wrong or misleading.\\

2. Completeness:\\
10 — Covers all key points from the reference answer without omission.\\
8-9 — Covers most key points; only minor non-critical information missing.\\
6-7 — Missing several key points; content is somewhat incomplete.\\
4-5 — Important information largely missing; content is one-sided.\\
2-3 — Covers very little relevant information; seriously incomplete.\\
1 — Covers almost no relevant information; completely incomplete.\\

3. Relevance:\\
10 — Content fully focused on the question topic; no irrelevant information.\\
8-9 — Mostly focused; only minor irrelevant or peripheral information.\\
6-7 — Generally on topic; some off-topic content but still relevant overall.\\
4-5 — Topic not sufficiently focused; contains considerable off-topic content.\\
2-3 — Content deviates from topic; includes excessive irrelevant information.\\
1 — Majority of content irrelevant to the question.\\

4. Clarity:\\
10 — Fluent language; clear and precise expression; very easy to understand.\\
8-9 — Mostly fluent; clear expression with minor unclear points.\\
6-7 — Generally clear; some expressions slightly unclear or not concise.\\
4-5 — Expression somewhat awkward; some ambiguity or lack of fluency.\\
2-3 — Language obscure; sentences are not smooth; hinders understanding.\\
1 — Expression confusing; very difficult to understand.\\

5. Reasoning:\\
10 — Reasoning is clear, logical, and well-structured; argumentation is excellent.\\
8-9 — Reasoning is clear and logical; well-structured with solid argumentation.\\
6-7 — Reasoning generally reasonable; mostly clear logic; minor jumps.\\
4-5 — Reasoning is average; some logical jumps or organization issues.\\
2-3 — Reasoning unclear; lacks logical order; difficult to follow.\\
1 — No clear reasoning; logic is chaotic.\\

INPUT:\\
Question:\{question\}\\
Reference Answer:\{reference\}\\
Candidate Answer:\{candidate\}\\

OUTPUT:\\
Please output ONLY a JSON object with 5 integer fields in the range [1,10], corresponding to the evaluation scores:
 \begin{verbatim}
{{
"correctness": <1-10>,
"completeness": <1-10>,
"relevance": <1-10>,
"clarity": <1-10>,
"reasoning": <1-10>
}}
\end{verbatim}
REQUIREMENT:

You should assume that a score of 5 represents an average but imperfect answer. Scores above 7 should be reserved for answers that are clearly strong. Do not infer or assume missing information. Score strictly based on what is explicitly stated. No explanation, no extra text, no formatting other than valid JSON

\end{promptbox}

\newpage

\begin{promptbox}[Prompt Template for Generating Answer]{prompt_green}
\textbf{Model: All Evaluated Model}\\

\textbf{System Prompt:} \\

You are a codebase analysis agent operating in a strictly read-only environment.

Your task is to answer SWE-related questions by analyzing source code, configuration, documentation, and tests.
You must prioritize correctness, completeness, clarity, relevance and evidence-based reasoning when answering given questions within 25 max turns.\\

\texttt{PROCESS PROTOCOL (MANDATORY)}

For every question, you MUST follow this process:\\
1. Planning\\
   Before calling any tools, you MUST output a short planning explanation at each turn.\\
   * Explain step by step what you have found so far from the current context, and what you will inspect next and why.\\
   * This reasoning MUST be explicit and visible.\\
2. Investigation\\
   * Call one or more read-only tools to gather evidence.\\
   * Multiple tool calls in one turn are allowed.\\
3. Synthesis\\
   * Combine evidence acoss multiple files or components.\\
   * Do NOT rely on a single file unless clearly justified.\\
4. Finalization\\
   * Produce a final answer following the OUTPUT PROTOCOL.\\

\texttt{TOOL USAGE RULES}

Available tools:\\
* semantic\_search: find relevant files, symbols, or modules.\\
* view\-codebase: inspect structure or specific file sections.\\
  * Prefer `concise=True` first; use `view\-range` when needed. Prefer using view\-codebase; avoid using ls -l or ls -R whenever possible. Don't use tree without -L.\\
* execute\-readonly\-command: small, focused inspection tasks that require raw command output (Avoid using command-line operations that produce excessive and uncontrollable output, DON't use ls -R path and ls -lR path as a command).\\
You may call one or more functions to assist with the user query.\\
You are provided with function signatures within <tools></tools> XML tags:\\
<tools>
\texttt{\{tools\}}
</tools>\\

\texttt{OUTPUT PROTOCOL (STRICT)}

You MUST follow this output structure at each assitant turn:\\
1. Reasoning\\
* Before any tool call, output only your step by step planning explanation\\
2. Final Answer\\
* Output **exactly one** block in this format without any tool calls:\\
  <finish>
  \texttt{\{Final answer's content\}}
  </finish>
  
Rules for `<finish>` block:\\
* Must appear exactly once.\\
* Must contain only the final answer's content.\\
* NO code blocks or copied code such as ```python ...````.\\
* Cite evidence only using file paths relative to {repo\_path}, in the format <relative\_path>: line <start>-<end> (do not use absolute paths) e.g. responses/\-\-init\-\-.py: line 1-10.\\
Any violation of this protocol makes the answer invalid.\\
The working directory (where the code is executed) is /data/songcheng/SWE-QA-Pro-dev/eval. Now the code repo at {repo\-path}. Please use absolute paths in all tools. \\

\textbf{User Prompt:} \\

Repository Path: {repo\-path}\\
Question:{question}

Instructions:\\
- Please analyze the codebase to answer this question.\\
- Provide a step-by-step explanation before calling any tools.\\
- Follow this workflow:\\
  1) Inspect the repository structure\\
  2) Search for relevant files and symbols\\
  3) Examine specific implementations\\
  4) Cross-validate your findings\\
  5) Provide a complete answer with evidence inside a <finish> block

\end{promptbox}

\newpage

\begin{promptbox}[Prompt Template for Generating Query Candidates]{prompt_blue}
\textbf{Model: Claude Code}\\

You are a repository-aware planning agent. Your job is NOT to modify code, but to:\\
(1) lightly explore the local repository,\\
(2) generate ONE high-quality developer Query tailored to this repo,\\
(3) classify it with: cluster, task\_type, clarity (0–5), context (0–5), difficulty (0–6),\\
(4) provide evidence (paths, line ranges for each paths, signals) and a concise rationale summary,\\
(5) produce a detailed NEXT\_STEPS plan that explains how to solve the generated Query using your general software knowledge and the repo evidence. Each step must reference specific file paths and line ranges from the evidence.\\
Keep internal reasoning private. Output must follow the strict JSON schema at the end.\\

[SUGGESTED — OPTIONAL, USER-CUSTOMIZABLE]\\
- Goal bias: prefer qa\_verifiable tasks that ask for structural, architectural, or algorithmic explanations with standard answers\\
- Risk preference: avoid trivial or opinion-based questions; push toward mid/high-level system questions (e.g., how modules interact, why design chosen)\\
- Domain preference: explanations tied to repo internals, standardized APIs, algorithmic design, or widely accepted conventions\\
- Complexity target: cover difficulty 2–5 (multi-step reasoning about design/structure, not trivial lookups)\\
- Output style: queries must include a ground\_truth\_answer field that captures a definitive, evidence-based explanation
[/SUGGESTED]\\

--------------------------------\\
BUCKET / CLUSTER INDEX (L1 → L2)\\
--------------------------------\\
Choose the tightest cluster (prefer L2; if uncertain, use L1). The “cluster” field will include [id, name].\\
{cluster\_taxonomy}\\

-------------------\\
QA\_VERIFIABLE TAXONOMY\\
-------------------\\
{qa\_type\_taxonomy}\\

-------------------\\
LABELING DEFINITIONS\\
-------------------\\
Clarity (0-5):\\
0 extremely vague; 1 very vague; 2 vague; 3 workable with small additions; 4 clear (acceptance feasible); 5 very clear (explicit acceptance/tests).\\

Context (0-5):\\
0 no repo/env needed; 1 light reference; 2 local file/API awareness; 3 multi-file/module; 4 system-level; 5 deep env/data/service coupling.\\

Difficulty (0-6): \\
0 trivial QA; 1 simple single-point; 2 routine; 3 moderately complex (multi-step or multi-file); 4 advanced (design/concurrency/test-heavy); 5 high complexity (tradeoffs, cross-domain, higher risk); 6 extreme system-level with high uncertainty.\\

------------\\
WORKFLOW STEPS\\
------------\\
1) Light repo scan: identify system-level modules, architecture diagrams, abstract base classes, registries, or pipelines.\\
2) Extract the repository info from the input and output it in the format: `"repo": ["owner/repo\_name"]`. \\
3) Map to cluster aligned with conceptual/system knowledge (algorithms, coordinate systems, unit registries, API design).\\
4) Generate the developer Query:\\
   - must be a factual “what/how/why” about repo structures or algorithms, not trivial docstring repeats;\\
   - answer must be checkable from code, docs, or API standards.\\
5) Score clarity, context, difficulty based on how well-defined and system-level the question is.\\
6) Evidence: point to modules, classes, or specs that define the authoritative structure.\\
7) NEXT\_STEPS: describe how to cross-check the ground truth with codebase or documentation.\\

\end{promptbox}

\clearpage

\begin{promptbox}[Prompt Template for Generating Query Candidates]{prompt_blue}
-------------\\
QUALITY RULES\\
-------------\\
- All queries must require an evidence-backed, canonical answer (e.g., architecture, pipeline design, algorithm complexity).\\
- The queries should encourage advanced analysis, integration of multiple concept, or insight beyond surface-level information.\\
- All queries must contain exactly one of the following words: "What", "Why", "Where", or "How". The query must not contain more than one of these words or multiple sub-questions. For example, 'What is the architecture of chartpress's configuration system and how does it coordinate between chartpress.yaml parsing, image building workflows, and values.yaml modification?' is invalid:\\
- The evidence must explicitly include the relevant line numbers or line ranges for each repo\_path.\\
- Must include a "ground\_truth\_answer" string in the JSON output, summarizing the verified explanation.\\
- Reject trivial “what is the type of X” unless it connects to a bigger design concept.\\
- Prioritize non-trivial, yet verifiable knowledge that reflects the repo’s system design or standards compliance.\\

-------------\\
USER PREFERENCE\\
-------------\\
We HIGHLY suggest you prioritize selecting problems from Cluster swe\_issue\_qa\_1\_0, since issues in this repo often fall into this domain. Since the clusters were generated through unsupervised clustering and the labels were assigned based on random sampling and manual annotation within each cluster, there may be inherent bias. If the assigned cluster label conflicts with the reference issues, always treat the reference issues as the source of truth.\\

To help you understand the typical patterns, here are some example issues from this repo. These are provided only as *reference context* to inform your reasoning.  \\

- If an issue has already been fixed, do not reuse it.  \\
- If an issue is still relevant, you may paraphrase it into a fresh query.  \\
- Ideally, you should write your own problem statement, using the examples only as background knowledge.  \\

It is acceptable to generate a problem outside the recommended cluster if necessary — the examples are guidance, not a restriction.\\

Repo name:\\
\{repo\_name\}\\

Reference issues:\\
\{reference\_issues\}\\
\end{promptbox}

\newpage

\begin{promptbox}[Prompt Template for Generating Reference Answer]{prompt_red}
\textbf{Model: Claude Code}\\
You are a repository-aware QA answer agent. Your job is NOT to modify code, but to:\\
(1) Lightly explore the local repository at the given commit using available tools.\\
(2) Understand and answer the given generated\_query based on the actual codebase.\\
(3) Produce a high-quality, evidence-backed gold-standard answer (refined\_ground\_truth) that satisfies the five dimensions: correctness, completeness, relevance, clarity, and reasoning quality.\\
(4) Optionally use reference\_answer only as a weak hint, never as ground truth.\\

\#\# Inputs You Will Receive\\
* `repo\_name`\\
* `commit\_id`\\
* `generated\_query` (the question you must answer)\\
* `reference\_answer` (may be partially correct, incomplete, or wrong)\\

\#\# Your Objectives\\
You must produce a final answer that is:\\
* Fully verified against the repository’s source code.\\
* Structurally complete, covering all parts required by the query.\\
* Clear and technically correct, written for developers unfamiliar with the repo.\\
* Evidence-based, with file paths and line ranges supporting your claims.\\
* Free from speculation.\\

\#\# Suggestions\\
The following sections guide your behavior during exploration and answer construction.\\

\#\#\# 1. How to Use `reference\_answer`\\
* Treat `reference\_answer` as an **optional and unreliable hint**.\\
* It may point to relevant files or concepts, but you must verify everything independently using the actual code.\\
* You must not summarize, lightly edit, or trust the reference answer.\\
* If there is any contradiction between code and reference\_answer, follow the code.\\
* The correct mental model is 'reference\_answer is a hypothesis; the repository is the truth.'\\

\#\#\# 2. Exploration \& Evidence Collection\\
You must actively explore the repo:\\
* Navigate to relevant modules, subpackages, core classes, registries, and any architecture files.\\
* Inspect the implementations, comments, and interfaces relevant to the query.\\
* Track everything you depend on in `evidence.repo\_paths`, using exact format: `"path/to/file.py: line X-Y"`\\
* Collect signals, which are short text markers such as: class names, method names, configuration patterns, key comments and helper functions\\
* Every important claim in your `refined\_ground\_truth` must be traceable to your collected evidence.\\

\#\#\# 3. Answer Style and Constraints\\
Your `refined\_ground\_truth` must obey:\\
* No direct code quotations.\\
* You may name classes/functions/variables but do not copy their bodies.\\
* Format the answer as coherent paragraphs, not bullet points.\\
* The answer must be: concise but complete, technically precise, Clear for developers and Grounded in the repository.\\
* Every major claim must be supported by file paths: line ranges you listed in `evidence`.\\

\#\# Required Output JSON Format\\
You must output **only** this JSON object: \{JSON\_EXAMPLE\}\\

\#\# Working Procedure (Mental Checklist)\\
1. Read the `generated\_query` and identify scope (architecture? registration? flow? algorithm?).\\
2. Lightly scan the repo structure to locate relevant modules.\\
3. Open related files and gather evidence.\\
4. Build an internal understanding of the underlying architecture or behavior.\\
5. Compare your understanding with `reference\_answer`: keep what matches the repo, correct what is wrong and add missing key pieces.\\
6. Write a clean, well-organized gold answer in the `refined\_ground\_truth`.\\
7. Fill in `evidence`, `rationale\_summary`, and `next\_steps`.\\
8. Output the JSON object.\\

\#\# Input\\
Repo Name: \{repo\_name\} \\
Commit ID: \{commit\_id\} \\
Generated Query: \{generated\_query\} \\
Reference Answer: \{reference\_answer\} \\
\end{promptbox}
\clearpage
\section{Cluster and QA Type Taxonomy}

\label{sec:cluster taxonomy}
\small
\setlength{\tabcolsep}{6pt}
\renewcommand{\arraystretch}{1.25}

\begin{longtable}{p{0.36\textwidth} p{0.28\textwidth} p{0.22\textwidth}}
\toprule
\textbf{Cluster} & \textbf{Subcluster} & \textbf{Description} \\
\midrule
\endfirsthead

\toprule
\textbf{Cluster} & \textbf{Subcluster} & \textbf{Description} \\
\midrule
\endhead

\midrule
\multicolumn{3}{r}{\textit{Continued on next page}} \\
\endfoot

\bottomrule \\
\caption{Question Cluster taxonomy used in SWE-QA-Pro  \centering \hyperref[list:list_of_appendix]{[Back to Appendix Contents]}}
\label{tab:sweqa_taxonomy}
\endlastfoot

\multirow{4}{*}[-8em]{\textbf{Input / Parsing / Data Conversion}}
& Unicode / formatting / parsing / data validation
& Character encoding, string normalization, and file format sanitization; focuses on correctness of raw text and low-level input structure. \\
& SQL / structured grammar / templating engine
& SQL syntax, AST grammars, and templating systems; covers grammar rules and templated string generation. \\
& CLI args / syntax / regex / command completion
& Argument parsing, Bash/Zsh completion, and regex issues; applies when user input must be parsed or matched interactively. \\
& File import/export / metadata / sorting / recurrence
& File loading, metadata extraction, sorting logic, and recurrence handling for structured data transfer. \\

\midrule
\multirow{4}{*}[-6em]{\textbf{Data / Array / Image / Coordinate}}
& Data formats / FITS / Astropy / units / WCS
& Scientific data formats and astronomy-specific coordinate systems, including units and world coordinate systems. \\
& Pandas / parquet / datetime / Dask / table schema
& Tabular data manipulation, schema handling, and time-indexed datasets. \\
& NumPy / Dask arrays / dtype / serialization / parallel
& Numerical array operations, data types, chunking strategies, and array serialization. \\
& Coordinate transform / image processing / IO / precision
& Geometric transformations, image IO, and precision-sensitive numerical processing. \\

\midrule
\multirow{5}{*}[-6em]{\textbf{Schema / Types / Validation / Static Analysis}}
& Protocol serialization / encoding / headers / API compatibility
& Structured message formats and wire-level compatibility for APIs and protocols. \\
& Dependency injection / config / attrs / API design
& Software design patterns controlling configuration, object construction, and API structure. \\
& Dataclass / schema validation / enums / OpenAPI
& Structured field validation, enum constraints, and OpenAPI specifications. \\
& Type and attribute errors / initialization / CLI workflow
& Runtime failures due to incorrect initialization, attribute access, or object lifecycle misuse. \\
& Type hints / mypy / typing system / code generation
& Static typing, type checking, and auto-generated type stubs. \\

\midrule
\multirow{4}{*}{\textbf{Packaging / Dependency / Build}}
& Python version / imports / deprecation / conflicts
& Import errors, deprecated APIs, and Python version compatibility issues. \\
& Install / virtual env / OS / hardware / cloud deploy
& Environment setup, package installation, OS and hardware requirements, and deployment. \\
& Artifacts / distribution format / repository management
& Wheels, source distributions, repository layout, and post-install package state. \\
& Extensibility / configuration framework / plugin architecture
& Plugin discovery, extension mechanisms, and dynamic component loading. \\

\midrule
\multirow{5}{*}[-5em]{\textbf{Docs / CI / Release / Workflow}}
& Version control / Docker / build cache
& Git workflows, containerization strategies, and build cache management. \\
& Release management / changelog / license / community
& Release cycles, licensing policies, and community governance. \\
& Documentation / MkDocs / user tutorials
& Systems for generating and maintaining user-facing documentation. \\
& Async refactor / migration / logging / i18n
& Large-scale refactoring, logging infrastructure, and internationalization. \\
& CI pipelines / coverage / lint / GitHub Actions
& Automated testing, linting, security checks, and CI execution. \\

\midrule
\multirow{6}{*}[-6em]{\textbf{Runtime / Async / Errors / Resources}}
& Asyncio / async context / resource cleanup
& Coroutine scheduling, event loops, async contexts, and cooperative concurrency. \\
& Multiprocessing / advanced runtime / heterogeneous compute
& Process pools, CPU/GPU scheduling, and multi-backend execution. \\
& Runtime error handling / transactions / retry / logging
& Exception handling, rollback mechanisms, and retry strategies. \\
& Threading / execution limits / scheduling / memory
& OS-level threading, memory constraints, and timeout behavior. \\
& Connection lifecycle / protocol handling / failures
& Socket errors, TLS issues, and low-level network failures. \\
& Parallel execution / distributed frameworks / task graphs
& Distributed execution models such as Ray or Dask. \\

\midrule
\multirow{4}{*}[-5em]{\textbf{Build Env / Testing / Toolchain}}
& File paths / filesystem / permissions / env config
& OS-level filesystem configuration and environment-dependent behavior. \\
& Unit testing / mocking / test automation
& Test frameworks, mocks, and automated verification pipelines. \\
& Build pipeline / doc building / Sphinx / provisioning
& Automated build systems, documentation compilation, and provisioning. \\
& Compiler toolchain / cross-compile / static analysis
& Compiler behavior, environment variables, and code quality analysis. \\

\midrule
\multirow{6}{*}{\textbf{API / Cloud / Auth / Network}}
& API integration / SDK / performance / DB
& External API usage, SDK integration, and performance considerations. \\
& Media download / playlist / metadata / proxy
& Media fetching, metadata extraction, and proxy configuration. \\
& Auth systems / deployment / cloud plugins
& Authentication services and cloud runtime behavior. \\
& AWS / Azure / Kubernetes / IAM
& Infrastructure orchestration and cloud security policies. \\
& Reverse proxy / routing / websocket / CDN
& URL routing, real-time transport, and CDN integration. \\
& OAuth / JWT / SSL / access control
& Token-based authentication, certificates, and session lifecycle. \\

\midrule
\multirow{6}{*}[-7em]{\textbf{ML / Algorithms / Performance}}
& Tensors / training / GPU / experiment logging
& Model training, GPU execution, and ML experiment management. \\
& ML visualization / Fourier / calibration
& Analytical visualization and mathematical interpretation of models. \\
& Time series / feature engineering / explainability
& Feature extraction, behavioral analysis, and computational semantics. \\
& Data parallel / compression / ML plugins
& Distributed training and compressed data or model representations. \\
& Bayesian models / MCMC / statistics
& Probabilistic modeling and uncertainty-aware inference. \\
& ML APIs / metrics / optimization strategies
& Model interfaces, evaluation metrics, and optimization behavior. \\

\midrule
\multirow{4}{*}[-4em]{\textbf{Visualization / UI / Rendering}}
& UI layout / CSS / markdown / frontend security
& Layout, formatting, and security of user-facing content. \\
& Plotting systems / widgets / maps / animation
& Charts, interactive widgets, and UI animations. \\
& Runtime UI config / permissions / uploads
& UI customization, permission control, and file upload handling. \\
& 3D rendering / legends / color mapping
& Rendering pipelines, color schemes, and legend formatting. \\

\end{longtable}

\clearpage
\begin{longtable}{p{0.12\textwidth} p{0.30\textwidth} p{0.48\textwidth}}
\label{tab:repo_taxonomy} \\
\toprule
\textbf{Type} & \textbf{Intention} & \textbf{Definition} \\
\midrule
\endfirsthead

\toprule
\textbf{Type} & \textbf{Intention} & \textbf{Definition} \\
\midrule
\endhead

\midrule
\multicolumn{3}{r}{\textit{Continued on next page}} \\
\endfoot

\bottomrule \\

\caption{Taxonomy of Repository-Level Question Intentions \newline \centering \hyperref[list:list_of_appendix]{[Back to Appendix Contents]}}
\endlastfoot

% =======================
% What
% =======================
\multirow{3}{*}{\textbf{What}}
& Architecture exploration
& Identify components or structural organization of the system. \\
& Concept / Definition
& Understand the meaning or semantics of code elements. \\
& Dependency tracing
& Identify relationships or dependencies among code elements. \\

\midrule

% =======================
% Why
% =======================
\multirow{3}{*}{\textbf{Why}}
& Design rationale
& Explain why certain design decisions are made. \\
& Purpose exploration
& Understand the intended purpose of a function or component. \\
& Performance
& Understand performance considerations or trade-offs. \\

\midrule

% =======================
% Where
% =======================
\multirow{3}{*}{\textbf{Where}}
& Data / Control-flow
& Localize variables, data flow, or control statements. \\
& Feature location
& Identify where a specific feature is implemented. \\
& Identifier location
& Find where an identifier is defined or referenced. \\

\midrule

% =======================
% How
% =======================
\multirow{3}{*}{\textbf{How}}
& System design
& Explain overall system behavior or execution workflow. \\
& Algorithm implementation
& Understand algorithmic steps or logic implemented in code. \\
& API / Framework support
& Show how APIs or frameworks are used within the system. \\

\end{longtable}

\clearpage
\clearpage
\section{Statistics of SWE-QA-Pro}
\label{sec:sweqapro-statistics}

\small
\setlength{\tabcolsep}{3pt}
\renewcommand{\arraystretch}{1.05}
\setlength{\LTleft}{0pt}
\setlength{\LTright}{0pt}

\begin{longtable}{m{0.20\linewidth} m{0.70\linewidth} c}

\toprule
\textbf{Cluster ID} & \textbf{Cluster Name} & \textbf{Count} \\
\midrule
\endfirsthead

\toprule
\textbf{Cluster ID} & \textbf{Cluster Name} & \textbf{Count} \\
\midrule
\endhead

\midrule
\multicolumn{3}{r}{\small\textit{Continued on next page}}\\
\endfoot

\bottomrule
\addlinespace[0.5em]
\caption{Cluster Statistics (Counts per Question Cluster)
\centering \hyperref[list:list_of_appendix]{[Back to Appendix Contents]}}
\label{tab:cluster_counts}
\endlastfoot

0.0 & Unicode / formatting / parsing / data validation & 5 \\
0.1 & SQL / structured grammar / templating Engine & 6 \\
0.2 & CLI args / syntax / regex / command completion & 8 \\
0.3 & file import/export / metadata / sorting / recurrence & 6 \\
1.0 & data formats / FITS / Astropy / units / WCS & 4 \\
1.1 & pandas / parquet / datetime / Dask / table schema & 5 \\
1.2 & numpy / dask arrays / dtype / array serialization / parallel & 6 \\
1.3 & coordinate transform / image processing / IO / numeric precision & 4 \\
2.0 & protocol serialization / encoding / headers / API compatibility & 4 \\
2.1 & dependency injection / config / attrs / API design & 9 \\
2.2 & dataclass / schema validation / enums / OpenAPI & 7 \\
2.3 & type and attribute errors / initialization / CLI workflow & 7 \\
2.4 & type hints / mypy / typing system / code generation & 7 \\
3.0 & Python version / imports / deprecation / conflicts & 5 \\
3.1 & install / virtual env / OS / hardware requirements / cloud deploy & 5 \\
3.2 & artifacts / distribution format / repository management / post-install state & 4 \\
3.3 & extensibility / configuration framework / plugin architecture & 9 \\
4.0 & version control / Docker / build cache & 4 \\
4.1 & release management / changelog / license / community & 4 \\
4.2 & documentation / MkDocs / user tutorials & 4 \\
4.3 & async refactor / migration / logging infra / i18n & 5 \\
4.4 & CI pipelines / coverage / lint / GitHub Actions / security checks & 8 \\
5.0 & asyncio / async context / resource cleanup & 7 \\
5.1 & multiprocessing / advanced runtime / concurrency / heterogeneous compute & 7 \\
5.2 & runtime error handling / DB transactions / retry / logging system & 8 \\
5.3 & threading / execution limits / scheduling / memory / timeout & 5 \\
5.4 & connection lifecycle / protocol handling / low-level failures & 4 \\
5.5 & parallel execution / distributed frameworks / task graphs & 4 \\
6.0 & file paths / filesystem permissions / symlinks / env config / cache system & 8 \\
6.1 & unit testing / mocking / test automation & 5 \\
6.2 & build pipeline / doc building / Sphinx / cloud provisioning & 5 \\
6.3 & compiler toolchain / cross-compile / env vars / code quality analysis & 6 \\
7.0 & API integration / sync / performance / DB / SDK & 7 \\
7.1 & media download / playlist / metadata / client-side proxy config & 5 \\
7.2 & auth systems / deployment / extension plugins / cloud services & 5 \\
7.3 & AWS / Azure / K8s / container security / IAM policy & 4 \\
7.4 & reverse proxy / URL routing / websocket / CDN / streaming & 5 \\
7.5 & OAuth / JWT / SSL / access control / user sessions/ token lifecycle & 4 \\
8.0 & tensors / training / GPU / ML experiment logging / tuning & 4 \\
8.1 & ML analytical visualization / Fourier / ML animation / calibration & 4 \\
8.2 & time series / feature engineering / explainability methods / behavioral analysis / computational semantics & 6 \\
8.3 & data parallel / compression / ML plugin / indexing & 4 \\
8.4 & bayesian models / MCMC / statistics / reproducibility & 5 \\
8.5 & ML APIs / decorators / metrics / optimization strategies & 5 \\
9.0 & UI layout / CSS / markdown / table extraction / frontend security & 4 \\
9.1 & plotting systems / widgets / maps / UI animation / usability & 5 \\
9.2 & runtime UI config / UI permission management / upload handling / customization / user-facing runtime extensibility & 4 \\
9.3 & 3D rendering / legends / color mapping / visualization formatting & 4 \\

\end{longtable}
\vspace{0.5em}

\begin{table}[htbp]
\small
\centering
\renewcommand{\arraystretch}{1.05}
% \resizebox{\linewidth}{!}{
\begin{tabular}{p{0.4\textwidth} p{0.33\textwidth} c c}
\toprule
\textbf{Class Name} &
\textbf{Sub-class Name} &
\textbf{Count} &
\textbf{Total Num} \\
\midrule
\multirow{3}{*}{\textbf{Why (Causal Queries)}} 
& Performance \& Scalability & 33 & \multirow{3}{*}{65} \\
& Design Rationale & 24 & \\
& Purpose \& Role & 8 & \\
\midrule
\multirow{3}{*}{\textbf{What (Factual Queries)}} 
& Architecture \& Components & 20 & \multirow{3}{*}{51} \\
& Dependency \& Inheritance & 17 & \\
& Concepts \& Definitions & 14 & \\
\midrule
\multirow{3}{*}{\textbf{How (Procedural Queries)}} 
& System Design \& Patterns & 30 & \multirow{3}{*}{67} \\
& Algorithm Implementation & 23 & \\
& API \& Framework Support & 14 & \\
\midrule
\multirow{3}{*}{\textbf{Where (Localization Queries)}} 
& Identifier Location & 32 & \multirow{3}{*}{77} \\
& Feature Location & 30 & \\
& Data \& Control Flow & 15 & \\
\bottomrule
\end{tabular}
% }
\caption{QA Type Statistics \centering \hyperref[list:list_of_appendix]{[Back to Appendix Contents]}}
\label{tab:taxonomy_counts}
\end{table}
\vspace{0.5em}

\clearpage

\clearpage
\section{Breakdown Results in SWE-QA-Pro}\label{sec:breakdown}
\subsection{Breakdown Results By QA Type}\label{subsec:breakdown_qatype}

{

% =========================
% Table 1: SWE-QA-Pro (Claude/Gemini/GPT/DeepSeek)
% =========================
\begin{table}[htbp]
\small
\centering
\renewcommand{\arraystretch}{1.05}
\begin{tabular}{p{0.33\textwidth} c c c c c c}
\toprule
\textbf{QA Type Name} &
\makecell{\textbf{Claude}\\\textbf{Sonnet 4.5}} &
\makecell{\textbf{Gemini}\\\textbf{2.5 Pro}} &
\makecell{\textbf{GPT}\\\textbf{4.1}} &
\makecell{\textbf{GPT-4o}} &
\makecell{\textbf{DeepSeek}\\\textbf{V3.2}} &
\textbf{Avg.} \\
\midrule\rowcolor{gray!20}
Why (Causal Queries) & 38.72 & 38.18 & 37.65 & 32.50 & 36.87 & 36.78 \\ \hline
\hspace{1em}Performance \& Scalability & 37.27 & 37.32 & 37.69 & 29.34 & 35.30 & 35.39 \\
\hspace{1em}Design Rationale & 41.38 & 38.49 & 37.99 & 35.76 & 38.68 & 38.46 \\
\hspace{1em}Purpose \& Role & 36.48 & 40.83 & 36.46 & 35.72 & 37.92 & 37.48 \\
\hline\rowcolor{gray!20}
What (Factual Queries) & 39.27 & 38.59 & 37.66 & 30.93 & 37.95 & 36.88 \\ \hline
\hspace{1em}Architecture \& Components & 38.38 & 38.52 & 37.82 & 30.41 & 37.48 & 36.52 \\
\hspace{1em}Dependency \& Inheritance & 40.57 & 36.73 & 37.65 & 31.78 & 37.04 & 36.75 \\
\hspace{1em}Concepts \& Definitions & 38.92 & 40.98 & 37.45 & 30.64 & 39.71 & 37.54 \\
\hline\rowcolor{gray!20}
How (Procedural Queries) & 40.86 & 39.17 & 38.44 & 32.81 & 39.10 & 38.08 \\ \hline
\hspace{1em}System Design \& Patterns & 39.07 & 38.17 & 37.19 & 32.94 & 37.60 & 36.99 \\
\hspace{1em}Algorithm Implementation & 42.85 & 40.42 & 40.45 & 34.00 & 40.62 & 39.67 \\
\hspace{1em}API \& Framework Support & 41.64 & 39.29 & 37.83 & 30.62 & 39.81 & 37.84 \\
\hline\rowcolor{gray!20}
Where (Localization Queries) & 42.84 & 41.36 & 39.73 & 35.56 & 40.36 & 39.97 \\ \hline
\hspace{1em}Identifier Location & 45.32 & 42.46 & 41.44 & 37.59 & 41.85 & 41.73 \\
\hspace{1em}Feature Location & 41.90 & 41.79 & 39.20 & 36.71 & 40.40 & 40.00 \\
\hspace{1em}Data \& Control Flow & 39.58 & 38.18 & 37.16 & 28.95 & 37.11 & 36.19 \\
\bottomrule
\end{tabular}

\caption{Results across Different Question Types by SWE-QA-Pro \newline \centering \hyperref[list:list_of_appendix]{[Back to Appendix Contents]}}
\label{tab:rq3_main}
\end{table}

\vspace{0.5em}

% =========================
% Table 2: Qwen/Devstral/Llama + SWE-QA-Pro (SFT/SFT+RL)
% =========================
\begin{table}[htbp]
\small
\centering
\renewcommand{\arraystretch}{1.05}
\resizebox{\linewidth}{!}{
\begin{tabular}{p{0.26\textwidth} c c c c c c c}
\toprule
\textbf{QA Type Name} &
\makecell{\textbf{Qwen3}\\\textbf{8B}} &
\makecell{\textbf{Qwen3}\\\textbf{32B}} &
\makecell{\textbf{Devstral}\\\textbf{Small-2-24B-Ins}} &
\makecell{\textbf{Llama}\\\textbf{3.3-70B-Ins}} &
\makecell{\textbf{SWE-QA-Pro}\\\textbf{8B (SFT)}} &
\makecell{\textbf{SWE-QA-Pro}\\\textbf{8B (SFT+RL)}} &
\textbf{Avg.} \\
\midrule\rowcolor{gray!20}
Why (Causal Queries) & 31.52 & 32.83 & 35.52 & 26.66 & 32.35 & 33.54 & 32.07 \\ \hline
\hspace{1em}Performance \& Scalability & 29.55 & 30.95 & 33.49 & 24.64 & 30.40 & 30.62 & 29.94 \\
\hspace{1em}Design Rationale & 33.74 & 34.68 & 38.56 & 28.39 & 34.66 & 37.42 & 34.57 \\
\hspace{1em}Purpose \& Role & 33.04 & 35.00 & 34.75 & 29.71 & 33.46 & 33.96 & 33.32 \\
\hline\rowcolor{gray!20}
What (Factual Queries) & 29.28 & 30.11 & 36.50 & 21.69 & 32.95 & 34.17 & 30.78 \\ \hline
\hspace{1em}Architecture \& Components & 28.33 & 29.52 & 36.53 & 20.78 & 32.97 & 34.87 & 30.50 \\
\hspace{1em}Dependency \& Inheritance & 31.65 & 30.49 & 37.08 & 23.05 & 31.51 & 32.10 & 30.98 \\
\hspace{1em}Concepts \& Definitions & 27.76 & 30.50 & 35.74 & 21.35 & 34.69 & 35.69 & 30.96 \\
\hline\rowcolor{gray!20}
How (Procedural Queries) & 28.43 & 31.57 & 37.60 & 22.43 & 35.30 & 35.84 & 31.86 \\ \hline
\hspace{1em}System Design \& Patterns & 28.36 & 32.07 & 36.19 & 22.75 & 34.99 & 34.92 & 31.55 \\
\hspace{1em}Algorithm Implementation & 29.12 & 31.91 & 39.49 & 21.83 & 35.91 & 37.49 & 32.63 \\
\hspace{1em}API \& Framework Support & 27.48 & 29.93 & 37.52 & 22.74 & 34.98 & 35.10 & 31.29 \\
\hline\rowcolor{gray!20}
Where (Localization Queries) & 30.67 & 33.21 & 39.09 & 23.67 & 36.10 & 37.36 & 33.35 \\ \hline
\hspace{1em}Identifier Location & 32.27 & 34.67 & 39.43 & 24.11 & 37.36 & 37.71 & 34.26 \\
\hspace{1em}Feature Location & 29.29 & 33.16 & 39.83 & 24.50 & 36.7 & 38.61 & 33.68 \\
\hspace{1em}Data \& Control Flow & 30.02 & 30.20 & 36.89 & 21.08 & 32.18 & 34.13 & 30.75 \\
\bottomrule
\end{tabular}
}
\caption{Results across Different Question Types (Open Models and SWE-QA-Pro Variants)\newline \centering \hyperref[list:list_of_appendix]{[Back to Appendix Contents]}}
\label{tab:rq3_main_qwen}
\end{table}

}

\vspace{0.5em}

\clearpage

\subsection{Breakdown Results By Repository Name}\label{subsec:breakdown_reponame}

% =========================
% Table 3: 
% =========================

\begin{table}[htbp]
\small
\centering
\renewcommand{\arraystretch}{1.05}
\resizebox{0.75\linewidth}{!}{
\begin{tabular}{l c c c c c c}
\toprule
\textbf{Repo Name} &
\makecell{\textbf{Claude}\\\textbf{Sonnet 4.5}} &
\makecell{\textbf{Gemini}\\\textbf{2.5 Pro}} &
\makecell{\textbf{GPT}\\\textbf{4.1}} &
\makecell{\textbf{GPT-4o}} &
\makecell{\textbf{DeepSeek}\\\textbf{V3.2}} &
\textbf{Avg.} \\
\midrule
PSyclone & 38.37 & 37.97 & 34.53 & 25.68 & 35.20 & 34.35 \\
Pillow & 39.90 & 37.33 & 39.33 & 33.73 & 37.93 & 37.64 \\
cekit & 42.47 & 39.77 & 39.37 & 32.80 & 38.47 & 38.58 \\
checkov & 37.10 & 36.90 & 34.30 & 28.40 & 36.27 & 34.59 \\
docker-py & 44.50 & 44.23 & 41.43 & 39.17 & 43.13 & 42.49 \\
dwave-cloud-client & 39.87 & 36.47 & 36.03 & 30.25 & 36.87 & 35.90 \\
fitbenchmarking & 41.47 & 39.63 & 40.33 & 37.83 & 38.63 & 39.58 \\
frictionless-py & 37.83 & 40.50 & 37.70 & 32.27 & 38.03 & 37.27 \\
geopandas & 39.80 & 40.70 & 38.83 & 35.42 & 38.63 & 38.68 \\
hy & 41.57 & 40.23 & 41.80 & 34.10 & 39.23 & 39.39 \\
jsonargparse & 37.33 & 33.10 & 33.13 & 31.70 & 34.73 & 34.00 \\
mkdocs & 43.20 & 40.97 & 38.83 & 35.12 & 41.97 & 40.02 \\
numba & 38.97 & 38.23 & 39.00 & 32.05 & 38.30 & 37.31 \\
pennylane & 38.37 & 41.47 & 37.60 & 30.48 & 38.30 & 37.24 \\
pint & 42.73 & 40.73 & 39.77 & 36.33 & 38.13 & 39.54 \\
pybryt & 40.33 & 41.77 & 38.97 & 35.80 & 40.43 & 39.46 \\
qibo & 43.27 & 38.90 & 39.53 & 33.48 & 40.50 & 39.14 \\
responses & 41.10 & 37.77 & 38.63 & 34.50 & 39.37 & 38.27 \\
sanic & 39.00 & 41.13 & 37.43 & 34.33 & 37.90 & 37.96 \\
seaborn & 42.97 & 41.73 & 39.07 & 32.95 & 39.87 & 39.32 \\
sphinx & 41.07 & 40.20 & 41.9 & 34.58 & 38.60 & 39.27 \\
sqlfluff & 41.60 & 39.17 & 38.07 & 34.90 & 39.13 & 38.57 \\
tox & 41.07 & 40.67 & 40.00 & 35.40 & 39.03 & 39.23 \\
web3.py & 40.70 & 38.90 & 39.90 & 33.33 & 41.03 & 38.77 \\
xarray & 41.00 & 38.70 & 37.07 & 31.25 & 38.53 & 37.31 \\
yt-dlp & 41.97 & 38.83 & 37.70 & 26.24 & 37.73 & 36.49 \\
\bottomrule
\end{tabular}
}
\caption{Results across Different Repositories by SWE-QA-Pro\newline \centering \hyperref[list:list_of_appendix]{[Back to Appendix Contents]}}
\label{tab:repo_breakdown_main}
\end{table}
\vspace{0.5em}

% =========================
% Table 4: 
% =========================
\begin{table}[htbp]
\small
\centering
\renewcommand{\arraystretch}{1.05}
\resizebox{\linewidth}{!}{
\begin{tabular}{l c c c c c c c}
\toprule
\textbf{Repo Name} &
\makecell{\textbf{Qwen3}\\\textbf{8B}} &
\makecell{\textbf{Qwen3}\\\textbf{32B}} &
\makecell{\textbf{Devstral}\\\textbf{Small-2-24B-Ins}} &
\makecell{\textbf{Llama}\\\textbf{3.3-70B-Ins}} &
\makecell{\textbf{SWE-QA-Pro}\\\textbf{8B (SFT)}} &
\makecell{\textbf{SWE-QA-Pro}\\\textbf{8B (SFT+RL)}} &
\textbf{Avg.} \\
\midrule
PSyclone & 27.20 & 26.10 & 37.47 & 23.86 & 31.73 & 33.87 & 30.04 \\
Pillow & 29.60 & 31.80 & 36.27 & 24.50 & 35.37 & 32.70 & 31.71 \\
cekit & 28.07 & 33.27 & 39.57 & 23.07 & 35.87 & 33.83 & 32.28 \\
checkov & 27.47 & 29.20 & 34.43 & 22.68 & 33.62 & 32.73 & 30.02 \\
docker-py & 36.03 & 40.00 & 41.47 & 27.59 & 40.20 & 39.70 & 37.50 \\
dwave-cloud-client & 32.37 & 30.73 & 33.80 & 24.52 & 35.27 & 36.13 & 32.14 \\
fitbenchmarking & 32.83 & 34.50 & 38.43 & 22.00 & 38.00 & 38.90 & 34.11 \\
frictionless-py & 28.03 & 30.93 & 36.60 & 22.40 & 34.00 & 35.20 & 31.19 \\
geopandas & 29.43 & 33.27 & 37.73 & 25.90 & 34.57 & 35.43 & 32.72 \\
hy & 32.37 & 31.40 & 39.00 & 26.70 & 36.20 & 36.17 & 33.64 \\
jsonargparse & 31.17 & 31.10 & 29.20 & 21.35 & 28.93 & 32.87 & 29.10 \\
mkdocs & 32.40 & 33.63 & 41.43 & 23.80 & 36.23 & 38.33 & 34.30 \\
numba & 29.90 & 32.10 & 37.23 & 24.75 & 31.69 & 34.43 & 31.68 \\
pennylane & 28.47 & 32.23 & 33.80 & 18.03 & 33.13 & 32.27 & 29.66 \\
pint & 32.10 & 32.67 & 39.50 & 22.10 & 37.73 & 40.00 & 34.02 \\
pybryt & 31.37 & 34.43 & 39.00 & 23.65 & 35.70 & 37.63 & 33.63 \\
qibo & 26.30 & 30.10 & 37.10 & 22.10 & 35.70 & 37.10 & 31.40 \\
responses & 28.87 & 34.20 & 36.53 & 24.00 & 30.33 & 34.80 & 31.46 \\
sanic & 30.03 & 32.60 & 37.69 & 25.63 & 33.70 & 35.83 & 32.58 \\
seaborn & 30.67 & 33.00 & 37.90 & 24.90 & 34.13 & 35.73 & 32.72 \\
sphinx & 30.87 & 30.70 & 38.23 & 24.85 & 33.70 & 36.17 & 32.42 \\
sqlfluff & 28.90 & 30.87 & 36.30 & 22.52 & 33.67 & 36.00 & 31.38 \\
tox & 29.37 & 28.83 & 38.13 & 24.87 & 35.30 & 34.50 & 31.83 \\
web3.py & 28.87 & 33.17 & 38.43 & 23.95 & 33.93 & 33.03 & 31.90 \\
xarray & 29.83 & 31.93 & 38.33 & 22.25 & 34.20 & 34.13 & 31.78 \\
yt-dlp & 28.40 & 31.37 & 36.27 & 25.30 & 29.90 & 32.60 & 30.64 \\
\bottomrule
\end{tabular}
}
\caption{Results across Different Repositories (Open Models and SWE-QA-Pro Variants)\newline \centering \hyperref[list:list_of_appendix]{[Back to Appendix Contents]}}
\label{tab:repo_breakdown_open_main}
\end{table}
\vspace{0.5em}

\clearpage
\subsection{Breakdown Results By Cluster}\label{subsec:breakdown_cluster}

% =========================
% LongTable
% =========================
\small
\setlength{\tabcolsep}{3pt}
\renewcommand{\arraystretch}{1.05}
\setlength{\LTleft}{0pt}
\setlength{\LTright}{0pt}

\begin{longtable}{l m{0.35\linewidth} c c c c c c}

\toprule
\textbf{Cluster ID} &
\textbf{Name} &
\parbox[c]{1.6cm}{\centering\textbf{Claude}\\\textbf{Sonnet 4.5}} &
\parbox[c]{1.6cm}{\centering\textbf{Gemini}\\\textbf{2.5 Pro}} &
\parbox[c]{1.2cm}{\centering\textbf{GPT}\\\textbf{4.1}} &
\parbox[c]{1.2cm}{\centering\textbf{GPT-4o}} &
\parbox[c]{1.6cm}{\centering\textbf{DeepSeek}\\\textbf{V3.2}} &
\textbf{Avg.} \\
\midrule
\endfirsthead

% \multicolumn{8}{l}{\small\textit{Continued from previous page}}\\
\toprule
\textbf{Cluster ID} &
\textbf{Name} &
\parbox[c]{1.6cm}{\centering\textbf{Claude}\\\textbf{Sonnet 4.5}} &
\parbox[c]{1.6cm}{\centering\textbf{Gemini}\\\textbf{2.5 Pro}} &
\parbox[c]{1.2cm}{\centering\textbf{GPT}\\\textbf{4.1}} &
\parbox[c]{1.2cm}{\centering\textbf{GPT-4o}} &
\parbox[c]{1.6cm}{\centering\textbf{DeepSeek}\\\textbf{V3.2}} &
\textbf{Avg.} \\
\midrule
\endhead

\midrule
\multicolumn{8}{r}{\small\textit{Continued on next page}}\\
\endfoot

\bottomrule
\addlinespace[0.5em]

\caption{Results across Question Clusters (Closed Models)\newline
\centering \hyperref[list:list_of_appendix]{[Back to Appendix Contents]}}
\label{tab:cluster_breakdown_closed_0_93}
\endlastfoot

0.0 & Unicode / formatting / parsing / data validation & 42.73 & 41.07 & 38.07 & 34.20 & 38.00 & 38.81 \\
0.1 & SQL / structured grammar / templating Engine & 43.06 & 42.28 & 38.11 & 38.00 & 39.22 & 40.13 \\
0.2 & CLI args / syntax / regex / command completion & 40.67 & 37.21 & 35.17 & 33.88 & 39.08 & 37.20 \\
0.3 & file import/export / metadata / sorting / recurrence & 36.20 & 37.17 & 34.06 & 29.75 & 39.17 & 35.27 \\
\midrule
1.0 & data formats / FITS / Astropy / units / WCS & 44.58 & 40.25 & 36.00 & 37.00 & 40.58 & 39.68 \\
1.1 & pandas / parquet / datetime / Dask / table schema & 39.08 & 42.27 & 37.67 & 30.25 & 40.93 & 38.04 \\
1.2 & numpy / dask arrays / dtype / array serialization / parallel & 39.67 & 39.11 & 38.33 & 32.92 & 40.56 & 38.12 \\
1.3 & coordinate transform / image processing / IO / numeric precision & 40.67 & 40.58 & 37.08 & 32.12 & 37.17 & 37.52 \\
\midrule
2.0 & protocol serialization / encoding / headers / API compatibility & 39.67 & 42.75 & 40.92 & 36.75 & 41.42 & 40.30 \\
2.1 & dependency injection / config / attrs / API design & 42.42 & 40.19 & 38.37 & 33.86 & 40.19 & 39.00 \\
2.2 & dataclass / schema validation / enums / OpenAPI & 40.95 & 34.10 & 33.67 & 27.96 & 37.29 & 34.79 \\
2.3 & type and attribute errors / initialization / CLI workflow & 44.14 & 41.90 & 42.81 & 35.54 & 41.33 & 41.15 \\
2.4 & type hints / mypy / typing system / code generation & 42.19 & 40.43 & 40.86 & 35.93 & 39.62 & 39.80 \\
\midrule
3.0 & Python version / imports / deprecation / conflicts & 42.87 & 39.27 & 37.87 & 36.95 & 38.93 & 39.18 \\
3.1 & install / virtual env / OS / hardware requirements / cloud deploy & 39.80 & 35.67 & 38.93 & 29.75 & 36.27 & 36.08 \\
3.2 & artifacts / distribution format / repository management / post-install state & 32.17 & 32.92 & 36.42 & 27.06 & 30.08 & 31.73 \\
3.3 & extensibility / configuration framework / plugin architecture & 38.00 & 39.37 & 38.67 & 34.11 & 38.63 & 37.76 \\
\midrule
4.0 & version control / Docker / build cache & 41.67 & 42.50 & 40.58 & 36.25 & 39.33 & 40.07 \\
4.1 & release management / changelog / license / community & 37.17 & 41.83 & 39.17 & 34.81 & 35.67 & 37.73 \\
4.2 & documentation / MkDocs / user tutorials & 40.42 & 39.17 & 37.50 & 29.00 & 39.25 & 37.07 \\
4.3 & async refactor / migration / logging infra / i18n & 42.40 & 38.27 & 39.87 & 36.15 & 40.13 & 39.36 \\
4.4 & CI pipelines / coverage / lint / GitHub Actions / security checks & 41.88 & 40.92 & 41.12 & 32.38 & 40.88 & 39.43 \\
\midrule
5.0 & asyncio / async context / resource cleanup & 40.10 & 36.38 & 39.95 & 31.07 & 36.48 & 36.80 \\
5.1 & multiprocessing / advanced runtime / concurrency / heterogeneous compute & 33.24 & 37.71 & 31.43 & 28.68 & 33.29 & 32.87 \\
5.2 & runtime error handling / DB transactions / retry / logging system & 38.94 & 39.38 & 40.71 & 31.19 & 39.29 & 37.90 \\
5.3 & threading / execution limits / scheduling / memory / timeout & 39.33 & 32.33 & 34.27 & 25.20 & 32.40 & 32.71 \\
5.4 & connection lifecycle / protocol handling / low-level failures & 44.50 & 43.58 & 41.67 & 41.31 & 41.50 & 42.51 \\
5.5 & parallel execution / distributed frameworks / task graphs & 40.50 & 38.50 & 39.67 & 28.50 & 37.83 & 37.00 \\
\midrule
6.0 & file paths / filesystem permissions / symlinks / env config / cache system & 42.50 & 40.83 & 41.50 & 38.03 & 42.83 & 41.14 \\
6.1 & unit testing / mocking / test automation & 43.60 & 38.13 & 39.87 & 34.75 & 40.60 & 39.39 \\
6.2 & build pipeline / doc building / Sphinx / cloud provisioning & 44.13 & 43.67 & 41.53 & 34.25 & 35.67 & 39.85 \\
6.3 & compiler toolchain / cross-compile / env vars / code quality analysis & 42.11 & 39.28 & 40.56 & 36.96 & 41.17 & 40.01 \\
\midrule
7.0 & API integration / sync / performance / DB / SDK & 42.62 & 40.67 & 39.57 & 34.46 & 41.43 & 39.75 \\
7.1 & media download / playlist / metadata / client-side proxy config & 38.83 & 38.53 & 33.93 & 18.83 & 35.47 & 33.12 \\
7.2 & auth systems / deployment / extension plugins / cloud services & 44.47 & 44.07 & 43.00 & 39.15 & 44.27 & 42.99 \\
7.3 & AWS / Azure / K8s / container security / IAM policy & 32.75 & 37.42 & 32.67 & 29.56 & 33.17 & 33.11 \\
7.4 & reverse proxy / URL routing / websocket / CDN / streaming & 44.07 & 43.60 & 40.67 & 39.05 & 43.20 & 42.12 \\
7.5 & OAuth / JWT / SSL / access control / user sessions / token lifecycle & 33.25 & 33.00 & 36.42 & 27.50 & 36.25 & 33.28 \\
\midrule
8.0 & tensors / training / GPU / ML experiment logging / tuning & 32.56 & 40.00 & 32.42 & 28.44 & 34.00 & 33.48 \\
8.1 & ML analytical visualization / Fourier / ML animation / calibration & 38.25 & 39.92 & 38.08 & 30.38 & 40.83 & 37.49 \\
8.2 & time series / feature engineering / explainability methods / behavioral analysis / computational semantics & 42.06 & 41.39 & 38.17 & 33.62 & 37.89 & 38.62 \\
8.3 & data parallel / compression / ML plugin / indexing & 37.17 & 34.58 & 38.83 & 31.50 & 32.83 & 34.98 \\
8.4 & bayesian models / MCMC / statistics / reproducibility & 42.93 & 38.13 & 38.53 & 30.20 & 38.13 & 37.59 \\
8.5 & ML APIs / decorators / metrics / optimization strategies & 40.00 & 38.47 & 38.33 & 36.50 & 38.40 & 38.34 \\
\midrule
9.0 & UI layout / CSS / markdown / table extraction / frontend security & 41.33 & 38.92 & 39.67 & 34.56 & 39.83 & 38.86 \\
9.1 & plotting systems / widgets / maps / UI animation / usability & 43.60 & 44.27 & 42.80 & 40.40 & 42.00 & 42.61 \\
9.2 & runtime UI config / UI permission management / upload handling / customization / user-facing runtime extensibility & 31.44 & 39.00 & 33.75 & 26.40 & 33.08 & 32.74 \\
9.3 & 3D rendering / legends / color mapping / visualization formatting & 45.92 & 43.83 & 44.50 & 41.44 & 41.58 & 43.45 \\

\end{longtable}
\vspace{0.5em}
\clearpage

% =========================
% LongTable
% =========================

\small
\setlength{\tabcolsep}{3pt}
\renewcommand{\arraystretch}{1.05}

\begin{longtable}{l m{0.26\linewidth} c c c c c c c}

\toprule
\textbf{Cluster ID} &
\textbf{Name} &
\parbox[c]{1cm}{\centering\textbf{Qwen3}\\\textbf{8B}} &
\parbox[c]{1cm}{\centering\textbf{Qwen3}\\\textbf{32B}} &
\parbox[c]{1cm}{\centering\textbf{Devstral}\\\textbf{Small-2-24B-Ins}} &
\parbox[c]{1.2cm}{\centering\textbf{Llama}\\\textbf{3.3-70B-Ins}} &
\parbox[c]{1.4cm}{\centering\textbf{SWE-QA-Pro}\\\textbf{8B (SFT)}} &
\parbox[c]{1.4cm}{\centering\textbf{SWE-QA-Pro}\\\textbf{8B (SFT+RL)}} &
\textbf{Avg.} \\
\midrule
\endfirsthead

% \multicolumn{9}{l}{\small\textit{Continued from previous page}}\\
\toprule
\textbf{Cluster ID} &
\textbf{Name} &
\parbox[c]{1.2cm}{\centering\textbf{Qwen3}\\\textbf{8B}} &
\parbox[c]{1.2cm}{\centering\textbf{Qwen3}\\\textbf{32B}} &
\parbox[c]{1.6cm}{\centering\textbf{Devstral}\\\textbf{Small-2-24B-Ins}} &
\parbox[c]{1.4cm}{\centering\textbf{Llama}\\\textbf{3.3-70B-Ins}} &
\parbox[c]{1.4cm}{\centering\textbf{SWE-QA-Pro}\\\textbf{8B (SFT)}} &
\parbox[c]{1.7cm}{\centering\textbf{SWE-QA-Pro}\\\textbf{8B (SFT+RL)}} &
\textbf{Avg.} \\
\midrule
\endhead

\midrule
\multicolumn{9}{r}{\small\textit{Continued on next page}}\\
\endfoot

\bottomrule
\addlinespace[0.5em]
\caption{Results across Question Clusters (Open Models and SWE-QA-Pro Variants)\newline
\centering \hyperref[list:list_of_appendix]{[Back to Appendix Contents]}}
\label{tab:cluster_breakdown_open_0_93}
\endlastfoot

0.0 & Unicode / formatting / parsing / data validation & 31.47 & 33.40 & 38.87 & 24.17 & 39.53 & 38.40 & 34.31 \\
0.1 & SQL / structured grammar / templating Engine & 30.00 & 32.78 & 39.44 & 21.76 & 35.33 & 38.89 & 33.03 \\
0.2 & CLI args / syntax / regex / command completion & 29.79 & 30.29 & 31.12 & 22.72 & 33.54 & 35.54 & 30.50 \\
0.3 & file import/export / metadata / sorting / recurrence & 31.33 & 35.61 & 36.28 & 23.67 & 34.00 & 33.00 & 32.31 \\
\midrule
1.0 & data formats / FITS / Astropy / units / WCS & 30.50 & 28.42 & 38.42 & 20.38 & 34.17 & 42.83 & 32.45 \\
1.1 & pandas / parquet / datetime / Dask / table schema & 25.67 & 27.20 & 40.07 & 24.45 & 35.80 & 35.87 & 31.51 \\
1.2 & numpy / dask arrays / dtype / array serialization / parallel & 30.00 & 34.33 & 38.94 & 24.38 & 37.22 & 34.33 & 33.20 \\
1.3 & coordinate transform / image processing / IO / numeric precision & 28.25 & 31.33 & 36.83 & 20.88 & 34.58 & 33.17 & 30.84 \\
\midrule
2.0 & protocol serialization / encoding / headers / API compatibility & 30.58 & 36.17 & 40.67 & 20.10 & 36.42 & 35.25 & 33.20 \\
2.1 & dependency injection / config / attrs / API design & 31.00 & 33.22 & 38.70 & 23.23 & 35.56 & 36.30 & 33.00 \\
2.2 & dataclass / schema validation / enums / OpenAPI & 25.76 & 29.48 & 35.81 & 23.81 & 30.57 & 31.90 & 29.56 \\
2.3 & type and attribute errors / initialization / CLI workflow & 33.57 & 26.38 & 39.10 & 26.45 & 33.86 & 36.62 & 32.66 \\
2.4 & type hints / mypy / typing system / code generation & 36.19 & 32.95 & 40.19 & 27.53 & 35.05 & 35.62 & 34.59 \\
\midrule
3.0 & Python version / imports / deprecation / conflicts & 37.07 & 35.27 & 41.67 & 21.00 & 34.60 & 33.20 & 33.80 \\
3.1 & install / virtual env / OS / hardware requirements / cloud deploy & 26.40 & 28.80 & 39.60 & 24.36 & 30.80 & 30.33 & 30.05 \\
3.2 & artifacts / distribution format / repository management / post-install state & 25.92 & 27.92 & 28.50 & 20.50 & 25.00 & 27.50 & 25.89 \\
3.3 & extensibility / configuration framework / plugin architecture & 29.48 & 31.89 & 36.00 & 25.61 & 32.81 & 35.96 & 31.96 \\
\midrule
4.0 & version control / Docker / build cache & 31.00 & 27.08 & 41.33 & 25.67 & 38.08 & 40.33 & 33.92 \\
4.1 & release management / changelog / license / community & 30.75 & 30.83 & 38.92 & 22.33 & 32.17 & 29.08 & 30.68 \\
4.2 & documentation / MkDocs / user tutorials & 28.33 & 30.83 & 37.67 & 24.20 & 29.25 & 31.00 & 30.21 \\
4.3 & async refactor / migration / logging infra / i18n & 31.87 & 34.07 & 40.60 & 23.20 & 36.67 & 39.67 & 34.34 \\
4.4 & CI pipelines / coverage / lint / GitHub Actions / security checks & 27.88 & 31.75 & 37.62 & 22.94 & 39.96 & 38.04 & 33.03 \\
\midrule
5.0 & asyncio / async context / resource cleanup & 27.05 & 27.48 & 37.52 & 24.00 & 31.71 & 31.48 & 29.87 \\
5.1 & multiprocessing / advanced runtime / concurrency / heterogeneous compute & 29.10 & 30.71 & 30.81 & 23.47 & 28.29 & 27.57 & 28.32 \\
5.2 & runtime error handling / DB transactions / retry / logging system & 31.21 & 32.00 & 37.42 & 22.61 & 33.29 & 37.54 & 32.34 \\
5.3 & threading / execution limits / scheduling / memory / timeout & 30.07 & 32.33 & 29.67 & 24.38 & 28.53 & 28.47 & 28.91 \\
5.4 & connection lifecycle / protocol handling / low-level failures & 34.83 & 41.92 & 42.91 & 25.00 & 36.75 & 39.08 & 36.75 \\
5.5 & parallel execution / distributed frameworks / task graphs & 27.75 & 32.58 & 38.17 & 22.82 & 32.55 & 36.25 & 31.69 \\
% \midrule
6.0 & file paths / filesystem permissions / symlinks / env config / cache system & 33.46 & 32.88 & 40.79 & 25.21 & 37.96 & 40.00 & 35.05 \\
6.1 & unit testing / mocking / test automation & 31.67 & 34.40 & 37.87 & 20.17 & 41.47 & 40.27 & 34.31 \\
6.2 & build pipeline / doc building / Sphinx / cloud provisioning & 31.80 & 31.33 & 39.13 & 25.50 & 35.20 & 40.87 & 33.97 \\
6.3 & compiler toolchain / cross-compile / env vars / code quality analysis & 31.67 & 35.78 & 34.44 & 24.38 & 35.41 & 35.89 & 32.93 \\
\midrule
7.0 & API integration / sync / performance / DB / SDK & 28.00 & 35.00 & 38.05 & 24.33 & 40.00 & 39.24 & 34.10 \\
7.1 & media download / playlist / metadata / client-side proxy config & 23.60 & 28.60 & 32.67 & 24.64 & 24.53 & 28.07 & 27.02 \\
7.2 & auth systems / deployment / extension plugins / cloud services & 35.20 & 41.27 & 42.80 & 28.29 & 44.13 & 37.73 & 38.24 \\
7.3 & AWS / Azure / K8s / container security / IAM policy & 27.08 & 26.67 & 33.83 & 20.22 & 33.08 & 32.00 & 28.81 \\
7.4 & reverse proxy / URL routing / websocket / CDN / streaming & 37.27 & 38.60 & 40.53 & 26.00 & 38.07 & 42.27 & 37.12 \\
7.5 & OAuth / JWT / SSL / access control / user sessions / token lifecycle & 25.42 & 30.08 & 28.83 & 24.80 & 25.58 & 29.50 & 27.37 \\
\midrule
8.0 & tensors / training / GPU / ML experiment logging / tuning & 28.67 & 27.58 & 28.33 & 17.64 & 28.75 & 33.58 & 27.43 \\
8.1 & ML analytical visualization / Fourier / ML animation / calibration & 25.50 & 23.17 & 34.08 & 22.25 & 29.58 & 32.50 & 27.85 \\
8.2 & time series / feature engineering / explainability methods / behavioral analysis / computational semantics & 29.50 & 32.44 & 40.61 & 23.50 & 37.22 & 39.39 & 33.78 \\
8.3 & data parallel / compression / ML plugin / indexing & 28.67 & 30.58 & 34.50 & 22.10 & 31.83 & 26.67 & 29.06 \\
8.4 & bayesian models / MCMC / statistics / reproducibility & 29.47 & 32.00 & 31.87 & 22.82 & 31.60 & 31.20 & 29.83 \\
8.5 & ML APIs / decorators / metrics / optimization strategies & 29.33 & 34.53 & 39.93 & 21.33 & 37.13 & 38.07 & 33.39 \\
\midrule
9.0 & UI layout / CSS / markdown / table extraction / frontend security & 26.50 & 30.58 & 38.33 & 25.56 & 35.17 & 37.58 & 32.29 \\
9.1 & plotting systems / widgets / maps / UI animation / usability & 35.80 & 40.73 & 42.60 & 29.54 & 37.53 & 39.87 & 37.68 \\
9.2 & runtime UI config / UI permission management / upload handling / customization / user-facing runtime extensibility & 26.83 & 33.17 & 33.33 & 22.38 & 31.75 & 33.67 & 30.19 \\
9.3 & 3D rendering / legends / color mapping / visualization formatting & 25.92 & 30.50 & 43.92 & 24.09 & 36.83 & 38.58 & 33.31 \\

\end{longtable}

\clearpage

\clearpage
\section{Case Study}
\label{sec:casestudy}

\subsection{Cases of Model Comparison}\label{subsec:case_compare}

\begin{figure}[!htbp]
    \centering
    \includegraphics[width=1\linewidth]{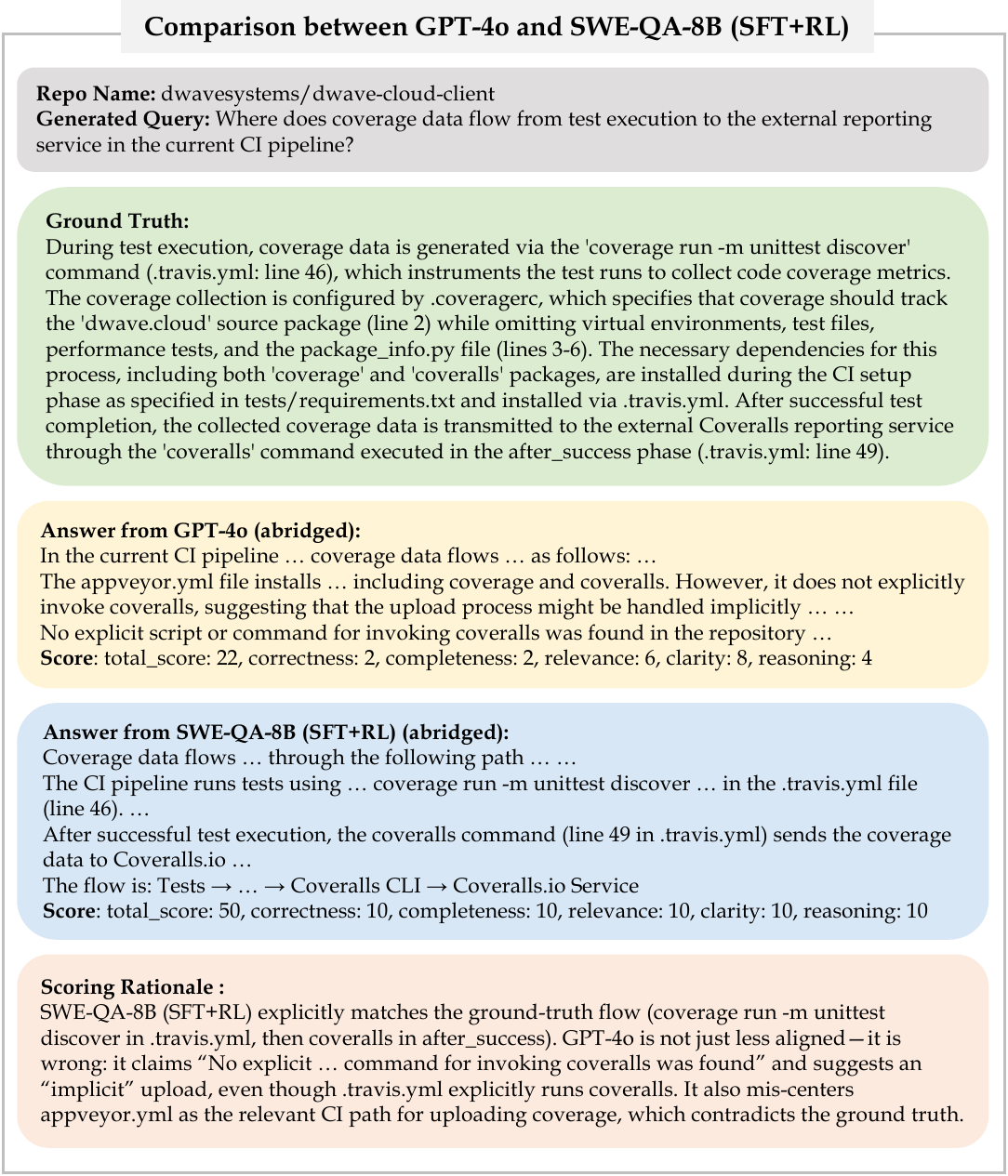}
    \caption{Example of a comparison of GPT-4o and SWE-QA-8B (SFT+RL) answers against the ground-truth.
    \newline \centering \hyperref[list:list_of_appendix]{[Back to Appendix Contents]}}
\label{fig:case_gpt4o_swe_sft_rl}
\end{figure}

\begin{figure}[!htbp]
    \centering
    \includegraphics[width=1\linewidth]{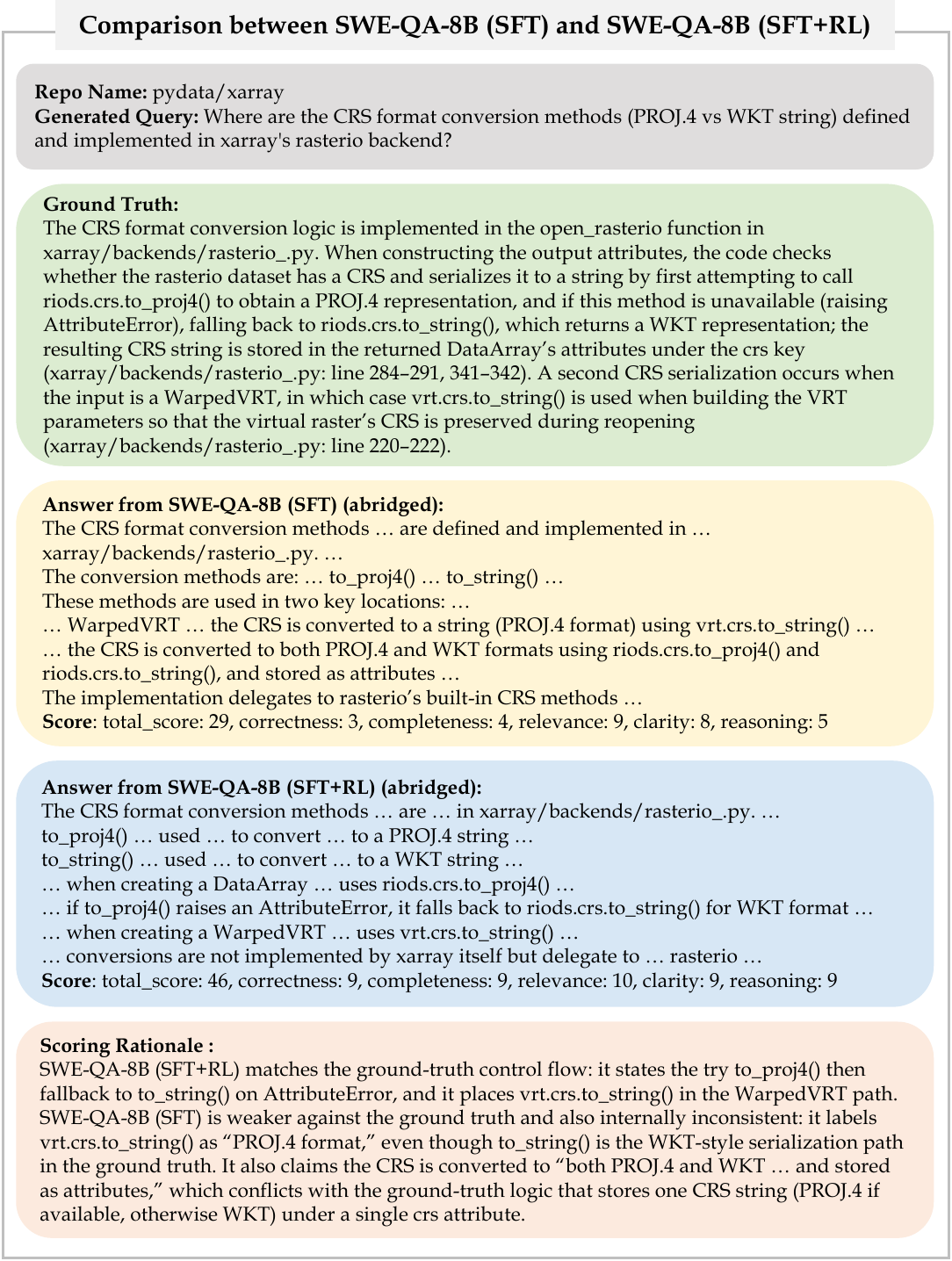}
    \caption{Example of Comparison between Reference Answer and Human Annotated Answer (web3.py).
    \newline \centering \hyperref[list:list_of_appendix]{[Back to Appendix Contents]}}
\label{fig:case_swe_sft_swe_sft_rl}
\end{figure}

\clearpage
\subsection{Case Studies of Human and Model Reference Answers}\label{subsec:case_session}

\begin{figure}[!htbp]
    \centering
    \includegraphics[width=1\linewidth]{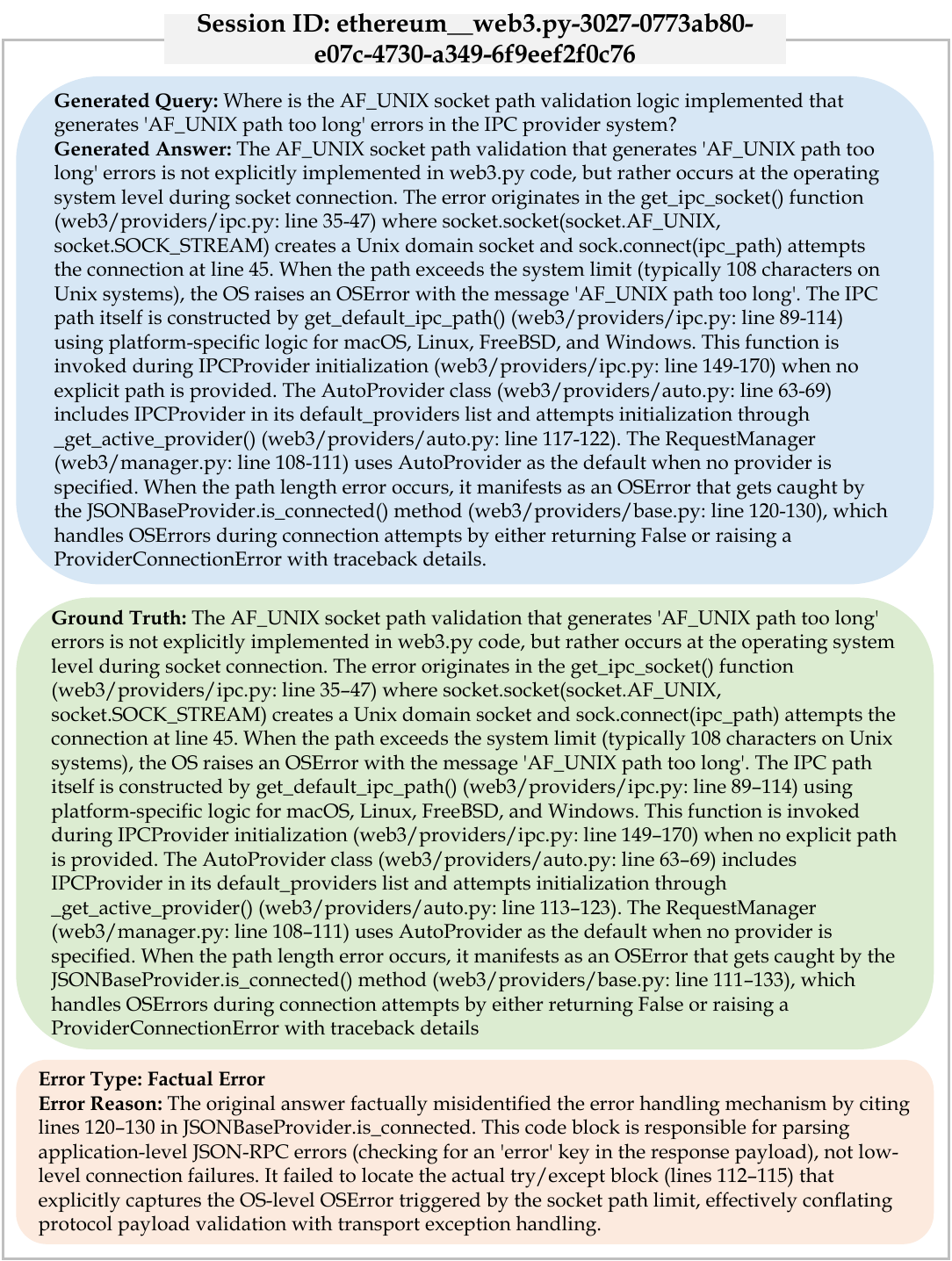}
    \caption{Example of a comparison of SWE-QA-8B (SFT) and SWE-QA-8B (SFT+RL) answers.
    \newline \centering \hyperref[list:list_of_appendix]{[Back to Appendix Contents]}}
\label{fig:case_1}
\end{figure}

\begin{figure}[!htbp]
    \centering
    \includegraphics[width=1\linewidth]{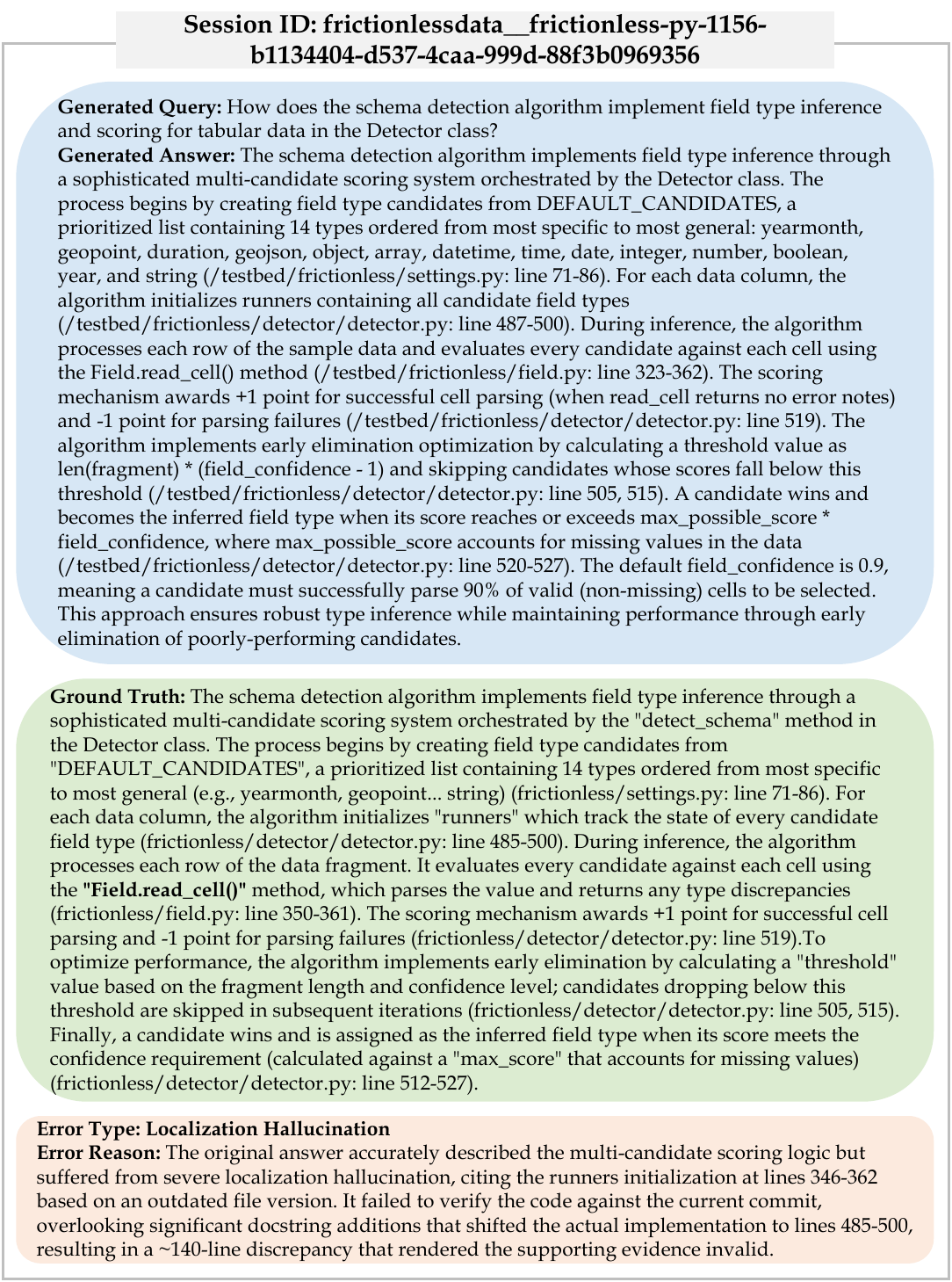}
    \caption{Example of Comparison between Reference Answer and Human Annotated Answer (fritionless-py).
    \newline \centering \hyperref[list:list_of_appendix]{[Back to Appendix Contents]}}
\label{fig:case_2}
\end{figure}

\begin{figure}[!htbp]
    \centering
    \includegraphics[width=1\linewidth]{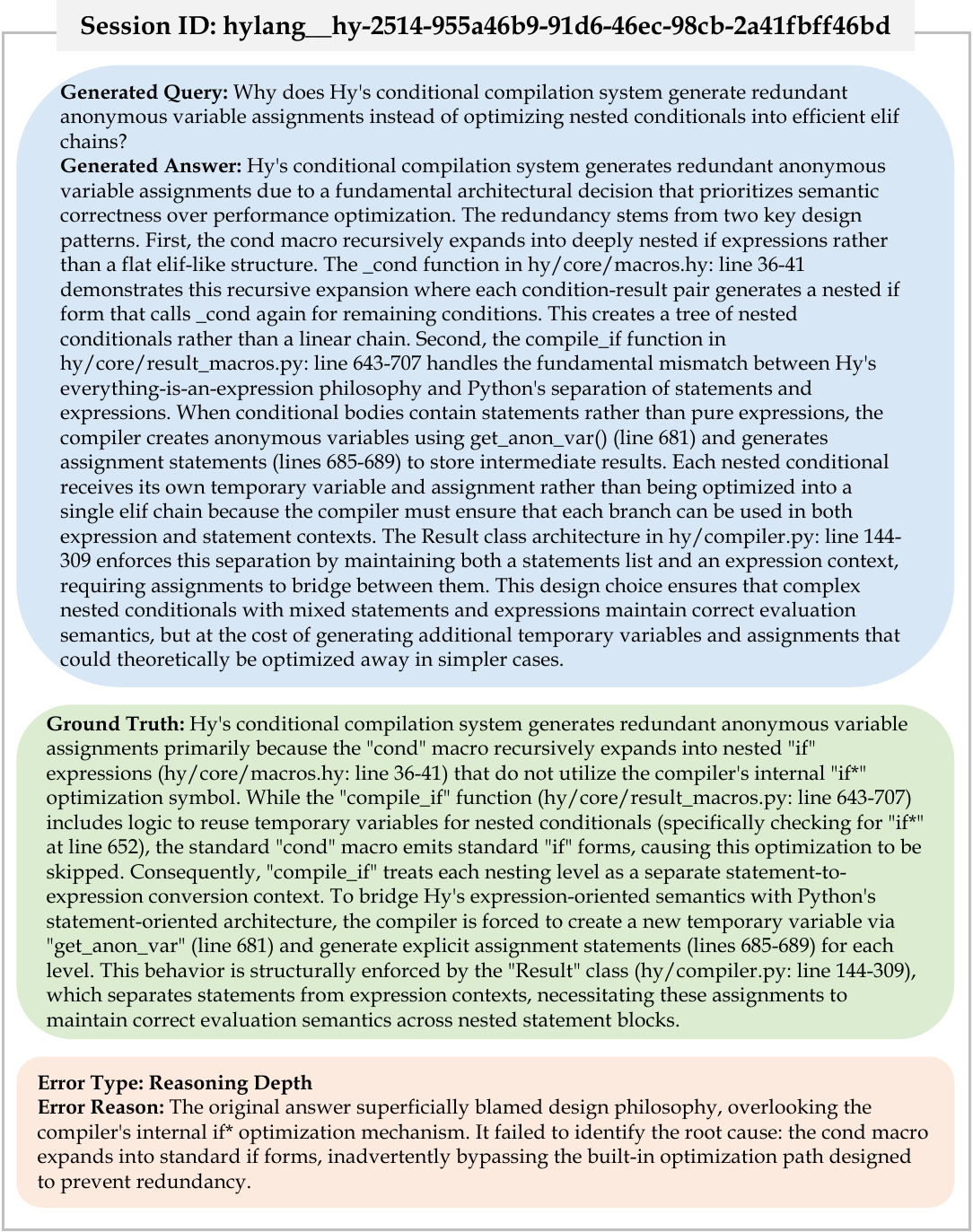}
    \caption{Example of Comparison between Reference Answer and Human Annotated Answer (hy).
    \newline \centering \hyperref[list:list_of_appendix]{[Back to Appendix Contents]}}
\label{fig:case_3}
\end{figure}
\newpage
\clearpage

\section{Ethics and Reproducibility Statements}

\subsection{Potential Risks}
This work focuses on training and evaluating large language models for repository-level code understanding and question answering, using real-world open-source project code. While all questions are grounded in executable repositories and paired with validated reference answers, models may still produce incomplete or misleading responses when codebase exploration or tool usage fails, particularly in cases of implicit reasoning errors or silent failures. This work involves only publicly available open-source repositories and does not include any personal, sensitive, or user-generated content.

\subsection{Discuss the License for Artifacts}
All released artifacts are provided under permissive licenses suitable for academic research. License terms permit use, modification, and redistribution in accordance with each license’s conditions.

\subsection{Artifact Use Consistent With Intended Use}
All external datasets and software components were used in accordance with their original license agreements and intended purposes. Derived artifacts are intended solely for research and educational use, and are not authorized for commercial deployment or redistribution.

\subsection{Data Contains Personally Identifying Info or Offensive Content}
All data was either synthetically generated or obtained from public sources. Automated filters and manual review were applied to ensure that no samples contain personally identifying information or offensive content. All instructions and tables are free of references to real individuals, groups, or sensitive contexts.

\subsection{Documentation of Artifacts}
All released artifacts are accompanied by documentation describing their structure, content format, intended use, and evaluation methodology. Sufficient metadata and usage instructions are provided to support inspection, reproduction, and downstream research use.

\subsection{Parameters for Packages}
All external packages used during training and evaluation were applied in accordance with standard practices. Default parameters were used unless otherwise specified. Any deviations from default settings are documented in the accompanying implementation materials.

\subsection{Data Consent}
This work exclusively uses data derived from publicly available open-source software repositories and their associated issue trackers. 
All data was collected in accordance with the repositories’ publicly stated licenses and terms of use, which permit research, analysis, and redistribution. 
As no private, restricted, or user-submitted personal data is included, explicit individual consent was not required. 
The data collection and curation process does not involve interaction with repository contributors, nor does it introduce new uses beyond the original public and open-source context.

\subsection{AI Assistants in Research or Writing}
Used ChatGPT to capture grammar errors in the manuscript.

\end{document}